%% file: Cecillon2024_revised2.tex
 \RequirePackage{booktabs}%
\documentclass[sn-mathphys-num]{sn-jnl}% Math and Physical Sciences Numbered Reference Style 
\usepackage{graphicx}%
\usepackage{multirow}%
\usepackage{amsmath,amssymb,amsfonts}%
\usepackage{amsthm}%
\usepackage{mathrsfs}%
\usepackage[title]{appendix}%
\usepackage{xcolor}%
\usepackage{textcomp}%
\usepackage{manyfoot}%
\usepackage{algorithm}%
\usepackage{algorithmicx}%
\usepackage{algpseudocode}%
\usepackage{listings}%
\usepackage{lmodern}
\usepackage{tikz}
\usetikzlibrary{arrows}
\usetikzlibrary{positioning,decorations.pathreplacing}
\tikzstyle{nn}=[circle,thick,draw=black!75,minimum size=6mm,fill=white]
\tikzstyle{rr}=[rectangle,rounded corners,thick,draw=black!75,minimum size=6mm,fill=white]

\theoremstyle{thmstyleone}%
\theoremstyle{thmstyletwo}%

\theoremstyle{thmstylethree}%

\definecolor{highlightCol}{RGB}{0, 0, 0}    % text color for highlights

\begin{document}

\title[Conversation-Based Multimodal Abuse Detection Through Text and Graph Embeddings]{Conversation-Based Multimodal Abuse Detection Through Text and Graph Embeddings}
% Detecting Abusive Messages through Multimodal Textual Content and Conversational Graph Representations
% "Multimodal Detection of Abusive Messages Through Text and Graph Embeddings"
% "Multimodal Abuse Detection by Representation Learning of Textual Content and Conversational Graphs"
% "Conversation-Based Multimodal Abuse Detection Through Text and Graph Embeddings"


\author*[1]{\fnm{Noé} \sur{Cécillon}}\email{noe.cecillon@univ-avignon.fr}

\author[1]{\fnm{Vincent} \sur{Labatut}}\email{vincent.labatut@univ-avignon.fr}

\author[2]{\fnm{Richard} \sur{Dufour}}\email{richard.dufour@univ-nantes.fr}

\affil*[1]{\orgdiv{LIA UPR 4128}, \orgname{Avignon Université}, \orgaddress{\country{France}}}

\affil[2]{\orgdiv{LS2N UMR 6004}, \orgname{Nantes Université}, \orgaddress{\country{France}}}


\abstract{Abusive behavior is common on online social networks, and forces the hosts of such platforms to find new solutions to address this problem. Various methods have been proposed to automate this task in the past decade. Most of them rely on the exchanged content, but ignore the structure and dynamics of the conversation, which could provide some relevant information.
In this article, we propose to use representation learning methods to automatically produce embeddings of this textual content and of the conversational graphs depicting message exchanges. While the latter could be enhanced by including additional information on top of the raw conversational structure, no method currently exists to learn whole-graph representations using simultaneously edge directions, weights, signs, and vertex attributes. We propose two such methods to fill this gap in the literature. We experiment with 5 textual and 13 graph embedding methods, and apply them to a dataset of online messages annotated for abuse detection. Our best results achieve an $F$-measure of 81.02 using text alone and 80.61 using graphs alone. We also combine both modalities of information (text and graphs) through three fusion strategies, and show that this strongly improves abuse detection performance, increasing the $F$-measure to 87.06. Finally, we identify which specific engineered features are captured by the embedding methods under consideration. These features have clear interpretations and help explain what information the representation learning methods deem discriminative.

\textcolor{red}{\textbf{Cite as:} Noé Cécillon, Vincent Labatut, and Richard Dufour. \textit{Conversation-Based Multimodal Abuse Detection Through Text and Graph Embeddings}. Springer Computing 107:124 (2025). DOI: \href{http://doi.org/10.1007/s00607-025-01463-6}{\texttt{10.1007/s00607-025-01463-6}}}
}

\keywords{Abuse detection, Text embeddings, Graph embeddings, Text classification}



\maketitle

%%%%%%%%%%%%%%%%%%%%%%%%%%%%%%%%%%%%%%%%%%%%%
\section{Introduction}
Detecting online abusive messages is a difficult task, as their authors often make use of innuendos, irony, implicit statements or even refer to past events shared by the persons involved in the conversation. Yet, such messages may have a lasting impact on the mental health, confidence, and sense of safety of the victims~\cite{Oreilly2018}. The standard approach to automatically detect online abusive messages relies on analyzing their content. Some markers of abuse, including swear words, and offensive or hateful expressions can easily be detected in this way. However, one can wonder if such expressions are necessarily representative of an abusive message. For instance, does quoting a previous message containing swear words make yours abusive? Does making a lame joke to a friend make you a harasser? The answer often depends on some general information, covering a wider scope than just the message itself. This is one of the reasons why it is difficult to automatically detect abusive comments. To overcome this, depending on the framework and application, multiple elements can be considered such as the shared history between users, demographic data, or the conversation preceding the problematic message.

The structure of a conversation can reflect the presence of an abusive author. Indeed, they tend to participate significantly more in the discussions than others and to receive more replies than regular users~\cite{Cheng2015}. Thus, the abusive author plays a central role in the conversation, and one can assume that the changes reflected in its structure can help discover an abuse case. Previous studies~\cite{Papegnies2017a, Papegnies2017b, Papegnies2019} show that modeling conversations under the form of graphs and characterizing them through various topological measures is efficient in detecting abusive messages. In certain cases, this structure-based approach which completely ignores the content of the messages can even achieve better results than a content-based method.

Be it to represent text or graphs, the standard feature-based approach used in~\cite{Papegnies2019} has some limitations, though. First, it requires manually searching the set of existing features and selecting the most appropriate ones to represent the data. To be comprehensive, such a search amounts to considering several hundred features, and the task may even require designing new \textit{ad hoc} features. Second, the computation of these features can be resource-intensive, in particular in the case of graphs~\cite{Papegnies2019}. Representation learning can solve this problem by automatically learning appropriate embedding representations of text and graphs. 
Lexical embeddings transform words into vectors that preserve their semantic and syntactic information. These methods offer robust representations that can help overcome the limitations of standard Natural Language Processing (NLP) techniques. Graph embedding methods are designed to learn representations of various parts of graphs (i.e. vertices, edges, subgraphs, or whole graphs). By construction, different methods are assumed to capture different aspects of the graph structure or properties.

%In this article, we want to answer the following research questions. RQ1: is it possible to use representation learning in place of feature engineering to perform abuse detection in conversations. RQ2: what is the discriminant information captured by these embedding methods. RQ3: is it possible to combine text- and graph-based embeddings.

In this article, we apply embedding techniques to the abuse detection task, in order to answer three Research Questions (RQ): 
\begin{itemize}
    \color{highlightCol}
    \item[RQ1] \textit{Is representation learning resulting in better abuse detection performance than feature engineered representations?}
    To answer this question, we apply a selection of embedding techniques to automatically learn representations of text messages and conversational graphs, and compare the resulting performance with that of the feature engineering-based approach from the literature~\cite{Papegnies2019}. 
    \item[RQ2] \textit{Are textual content and conversation structure complementary when detecting abuse?} 
    This question is based on the observation that combining multiple information modalities to analyze textual documents has proved effective on multiple tasks including sentiment analysis~\cite{Gandhi2023}, named entity recognition~\cite{Wu2020}, and recommendation~\cite{Huang2019}. For abuse detection, existing multimodal methods improve the performance of traditional text-oriented techniques. However, they are mainly centered around the combination of content and contextual data. The structure of the conversation is almost always overlooked in the literature. This is certainly related to the fact that practically all available corpora are constituted of \textit{independent messages}, and not \textit{full conversations}. To answer this question, we build a representation that takes advantage of both modalities (textual content and conversational structure), and study the effect on our abuse detection task. 
    \item[RQ3] \textit{Which discriminant information is automatically captured (or missed) by the embeddings, compared to the manually engineered features?} 
    We answer this question by studying how adding particularly discriminant features to embeddings impacts their abuse detection performance.
\end{itemize}

\textcolor{highlightCol}{We formulate the abuse detection task as a binary classification problem, where one wants to determine, for a given message, if it falls into the \textit{Abusive} or \textit{Non-abusive} classes. Our experimental setup consists of two main steps. First, we compute a vector-based representation of the message and its context, derived either from textual content or the conversational graph. Second, we use these representations to train a classifier to distinguish between abusive and non-abusive messages.} 

We identify four main contributions in our work: 
\begin{enumerate}
    \color{highlightCol}
    \item We fill a gap in the literature by proposing two whole-graph embedding approaches that simultaneously take into account edge weights, signs, directions, and vertex attributes. 
    \item We conduct extensive experiments to compare textual and graph embedding methods for the abuse detection task, including the two proposed whole-graph embedding approaches, thereby answering RQ1.
    \item We combine textual and graph-based methods through three fusion strategies to estimate the complementarity of these modalities and answer RQ2. 
    \item Since the learned representations are not directly interpretable, it is not straightforward to understand exactly which information is captured by the embeddings. Therefore, we also perform an in-depth analysis of the methods to detect which standard text and graph features are captured by which methods, allowing us to answer RQ3.
\end{enumerate}

The rest of this article is organized as follows. In Section~\ref{sec:EmbeddingMethods}, we present the existing text and graph embedding methods that we use in our experiments. In Section~\ref{sec:ProposedMethods}, we propose two whole-graph embedding methods able to take into account simultaneously edge weights, signs, directions, and vertex attributes. Then, we present our experimental setup, including the dataset, in Section~\ref{sec:ExperimentalSetup}. We put all these methods into practice on our abuse detection task and discuss their results in Section~\ref{sec:ClassificationResults}. Then, we focus on determining which information is captured by every method in Section~\ref{sec:EmbeddingFeatureStudy}. Finally, we review our main findings in Section~\ref{sec:Embeddingconclusion}, and identify some perspectives for this work.

%%%%%%%%%%%%%%%%%%%%%%%%%%%%%%%%%%%%%%%%%%%%%
\section{Selection of Existing Embedding Methods}
\label{sec:EmbeddingMethods}
In this section, we describe the state-of-the-art embedding methods that we select to experiment on our abuse detection task. We distinguish between the lexical methods (Section~\ref{sec:EmbLexicalMethods}), which we apply to the textual content of the messages exchanged among the users, and the graph-based methods (Section~\ref{sec:EmbGraphMethods}), which we use to represent the conversational networks modeling inter-user interactions.

%%%%%%%%%%%%%%%%%%%
\subsection{Text Embeddings}
\label{sec:EmbLexicalMethods}
Representation learning for textual data consists in transforming textual inputs into numeric vectors, which can then be used by machine learning algorithms for various tasks. Word embedding methods learn dense vectors of fixed size that represent a set of words, one vector per token in the vocabulary. They efficiently encode the semantic and syntactic information of words: those that are used in similar ways obtain very close representations in the vector space, naturally capturing their meaning. However, representing polysemy and homonymy is still challenging. One can distinguish two categories of word embedding approaches: the ones that are context-invariant and output a \textit{fixed} representation for each word, and the \textit{contextualized} ones that can generate different embeddings for a given word depending on the context in which it is used. 

In the abuse detection literature, the tools obtaining the best performance are typically text embedding methods pre-trained or fine-tuned over corpora of abusive messages~\cite{Cecillon2024phd}. A good example is HateBERT~\cite{Caselli2021}, which is a version of BERT trained on a large collection of offensive Reddit messages. However, one characteristic of our dataset is that it consists of French messages, and therefore, these tools cannot be directly applied to our case without a significant drop in performance. For this reason, in order to conduct our experiments, we select five standard methods pre-trained for the French language, which we fine-tune on our own corpus. The first two are context-invariant while the other three are context-sensitive. 

%%%
\medskip\noindent\textbf{Word2vec}~\cite{Bojanowski2017} W2V generates fixed representations of words. It proposes the SkipGram model architecture, which has been reused in numerous methods since, even outside the text processing domain. The model learns representations while trying to preserve the semantic and/or syntactic similarity of words. Word2vec is based on a distributional hypothesis and learns the meaning of words from a large corpus of texts. Technically, it uses a neural network that has an input layer, an output layer, and a projection layer. The latter constitutes the word embedding. The output layer is used to perform a classification task allowing to train of the model.

%%%
\medskip\noindent\textbf{fastText}~\cite{Bojanowski2017} FT is an extension of Word2vec that was developed to improve the representations of uncommon words. It breaks words down to $N$-grams of characters instead of treating full words as Word2vec does. In this way, each word can be represented as characters $N$-grams (with $3 \leq $N$ \leq 6$, generally). Once the character $N$-grams are extracted, a SkipGram model learns their representations. The representation of a word is obtained by summing all its $N$-gram representations. With this approach, rare and out-of-vocabulary words can still get a quality representation, since it is very likely that at least a portion of their $N$-grams appear in other words. 

%%%
\medskip\noindent\textbf{CamemBERT}~\cite{Martin2020en} \textbf{\& FlauBERT}~\cite{Le2020en} These are two French word embedding models directly adapted from the RoBERTa~\cite{Liu2019} architecture. Therefore, they share a lot of similarities. The main difference between them lies in the way that they mask tokens in the masked language model. CamemBERT also uses more training data than FlauBERT. They both provide contextualized word representations. 

%%%
\medskip\noindent\textbf{Flair}~\cite{Akbik2018} Unlike the four previous approaches, Flair learns representations that are built at the character level and not the word level. This approach allows Flair to be particularly efficient when dealing with rare or misspelled words. This architecture uses a bi-directional LSTM operating on characters. The model is trained to predict the next character for each element in the sequence to process. Therefore, it learns two hidden representations for each character of the sequence: one for the forward network and one for the backward network. The embedding of a word is obtained by combining the forward representations of the characters located before the end of the word and the backward representations of the characters located after the beginning of the word.

%%%%%%%%%%%%%%%%%%%
\subsection{Graph Embeddings}
\label{sec:EmbGraphMethods}
In this article, we want to leverage conversational networks to handle our abuse detection task. We come back to the graph extraction method in further detail in Section~\ref{subsec:Dataset}, but we need to summarize it here first, before describing the graph embedding methods. Each graph represents a single conversation, i.e. a sequence of messages centered around a message of interest. This so-called \textit{targeted message} is the one we want to classify, the rest of the messages constitute its context. In this graph, vertices model users participating in the conversation, and are described by certain individual attributes. Edges represent the exchanges of messages between these users, and are characterized by a direction, a weight, and a sign. 

Based on this graph model, the abuse detection task consists in classifying conversational networks. \textcolor{highlightCol}{To do so, the ideal representation learning approach is to use a whole-graph embedding method able to support all the available information: graph structure, vertex attributes, edge directions, weights and signs. However, such a method does not exist in the literature~\cite{Cecillon2024phd, Cecillon2024}.} To circumvent this problem, we adopt two strategies. The first is to use a set of widespread existing methods able to handle at least \textit{a part} of the available information. We select them so that they represent a diversity of approaches and focus on preserving various aspects of the graph. In the rest of this section, we briefly describe these methods. Approximately half of them treat the graph at the \textit{vertex} scale (Section~\ref{sec:EmbGraphMethodsVtx}), while the rest handle the graph as a \textit{whole} (Section~\ref{sec:EmbGraphMethodsGr}). The second strategy consists in extending existing graph embedding methods so that they take advantage of all the available information (this is the object of Section~\ref{sec:ProposedMethods}).

%%%%%%%%%
\subsubsection{Vertex Embedding Methods}
\label{sec:EmbGraphMethodsVtx}
In the conversational graphs used to perform the abuse detection task, all vertices are not equal. One of them represents the author of the targeted message, which we call the \textit{targeted vertex}. We assume that this vertex plays a particular role if an abuse is occurring at this moment of the conversation. We can suppose that this vertex is the most important in the graph, and that characterizing it individually could be enough to represent the full conversation. The vertex embedding methods presented in this section allow learning the representation of this vertex. As an alternative, we also experimented with averaging the representations of all the vertices in the graph. However, the obtained performance is systematically lower, so we only focus on the targeted vertex-based representation in the rest of the article. 

%%%
\medskip\noindent\textbf{DeepWalk}~\cite{Perozzi2014} DW samples vertex sequences using uniform random walks and then applies the standard \textit{SkipGram} model~\cite{Mikolov2013} to generate the representations. This model takes a vertex as input and aims at predicting its context, i.e. the vertices in its neighborhood. With this method, vertices with similar contexts share similar representations.

%%%
\medskip\noindent\textbf{Node2vec}~\cite{Grover2016} N2V was developed following the idea of \textit{DeepWalk}. The main difference is that \textit{Node2vec} uses \textit{biased} random walks to provide a more flexible notion of a vertex's neighborhood and better integrate the notion of structural equivalence. More specifically, it uses 2 parameters to bias the transition probabilities between vertices. The \textit{return parameter} controls the likelihood of immediately revisiting a vertex in the random walk. The \textit{in-out parameter} can restrict the walks to a local neighborhood or conversely, increase the probability of visiting vertices that are further away from the current one.

%%%
\medskip\noindent\textbf{Walklets}~\cite{Perozzi2017} WL is an extension of DeepWalk which aims at explicitly modeling multi-scale relationships, i.e. combining distinct views of vertex relationships at different granularity levels. The key change is in the random walk sampling algorithm, as the walks can now skip some vertices to reach farther parts of the network. It creates a representation for each size of \textit{skip}) and the output representation is the result of their concatenation.

%%%
\medskip\noindent\textbf{BoostNE}~\cite{Li2019} BNE is a multi-level graph embedding framework that learns multiple graph representations at different granularity levels. Inspired by boosting, it is built on the assumption that multiple weak embeddings can lead to a stronger and more effective one. It applies an iterative process to a closed-form vertex connectivity matrix. This process successively factorizes the residual obtained from the previous factorization, to generate increasingly finer representations. The sequence of representations produced is then assembled to create the final embedding.

%%%
\medskip\noindent\textbf{GraphWave}~\cite{Donnat2018} GW mimics a physical process of propagating some energy through the graph structure, starting from the vertex of interest. The way this energy is diffused over the graph is assumed to characterize the vertex and its neighborhood. The heat wavelet diffusion coefficients thus leveraged are used as a probability distribution and embedded into a representation space by calculating the characteristic function for each vertex coefficient.

%%%
\medskip\noindent\textbf{$k$-hop Graph Neural Networks}~\cite{Nikolentzos2020} KH-GNNs identify a fundamental limitation in GNNs and propose a more expressive architecture to solve this problem. Instead of updating a vertex's representation by aggregating information from its direct neighbors, this method aggregates its direct neighbors and its $k$-hop neighborhood through a graph neural network. This architecture is capable of identifying graph properties that are not captured by standard GNNs.

%%%%%%%%%
\subsubsection{Whole-Graph Embedding Methods}
\label{sec:EmbGraphMethodsGr}
Using a description of the whole graph amounts to considering the entire conversation at once when performing the classification. It can help better capture the global dynamics of the exchanges between users in the graph.

%%%
\medskip\noindent\textbf{Family of Graph Spectral Distances}~\cite{Verma2017} FGSD is a conceptually simple yet powerful graph representation method. The learned representations exhibit certain uniqueness, stability, and sparsity properties while being fast to compute. FGSD is built on the assumption that the graph atomic structure is encoded in the multiset of all vertex pairwise distances. It computes the Moore-Penrose Pseudoinverse spectrum of the graph Laplacian. A vector representation of the whole graph is constructed from the histogram of this spectrum.

%%%
\medskip\noindent\textbf{Spectral Features}~\cite{deLara2018} SF is a simple and fast algorithm that computes the spectrum of the normalized graph Laplacian, keeping only the $k$ smallest positive Eigenvalues. These Eigenvalues, in ascending order, form the representation of the graph. If the graph has fewer than $k$ vertices, the representations are right-padded with zeros. This simple method was originally developed as a baseline for graph classification but shows competitive results.

%%%
\medskip\noindent\textbf{Nested Graph Neural Networks}~\cite{Zhang2021} NGNN can be seen as a two-level GNN. It first learns a representation of each vertex that encodes the general local subgraph information around it before aggregating them. To obtain a vertex representation, this method first extracts a rooted subgraph and uses an arbitrary GNN to learn an intermediate representation for every vertex of the subgraph. The intermediate vertex representations are then pooled to obtain the subgraph representation which is used as the final representation of the root vertex. These representations are, in turn, pooled into a single one to represent the whole graph.

%%%
\medskip\noindent\textbf{Graphormer}~\cite{Ying2021} GO is built upon the standard Transformer architecture, originally designed for NLP~\cite{Vaswani2017}. Instead of processing the tokens that form a text, GO handles the vertices constituting a graph. In order to capture the graph structure, it uses two types of encodings in place of the standard positional encoding: vertex importance is represented through a \textit{Centrality Encoding}, and relations between pairs of vertices through a \textit{Spatial Encoding}. Akin to BERT~\cite{Devlin2019} for text, which uses a special \texttt{[CLS]} token to produce whole sentence representations, GO includes a special vertex called \texttt{[VNode]} to obtain whole-graph representations. This \textit{virtual node} is connected to all the other vertices in the graph, and is similar to the master node used by WSGCN (cf. below).

%%%
\medskip\noindent\textbf{Graph2vec}~\cite{Narayanan2017} G2V considers a graph as a collection of subgraphs, which are then used to learn a representation of the whole graph. More precisely, it looks for the so-called \textit{rooted} subgraph, i.e. vertex neighborhoods of a certain order. G2V extracts a \textit{rooted} subgraph around every vertex in the graph following the Weisfeiler--Lehman relabeling process~\cite{Weisfeiler1968}. At the end of this process, 2 isomorphic rooted subgraphs should get the same label. The extracted subgraphs are then used to train a SkipGram model.

%%%
\medskip\noindent\textbf{Signed Graph2vec}~\cite{Cecillon2024} SG2V is an adaptation of the original Graph2vec approach that can learn representation of signed graphs. To take advantage of this additional information, SG2V uses two variants of the Weisfeiler--Lehman relabeling procedure able to handle edge signs. In particular, SG2V includes edge signs in the generated labels, and uses the positive and negative degrees as initial labels. The first variant, \texttt{SG2Vsb}, is based on the notion of \textit{Structural Balance}~\cite{Heider1946, Harary1953}, whereas the second, \texttt{SG2Vgb} relies on a generalization of this concept. 

%%%
\medskip\noindent\textbf{Whole SGCN}~\cite{Cecillon2024} WSGCN is another signed whole-graph embedding method. It is based on Signed Graph Convolutional Networks (SGCN)~\cite{Derr2018}, a signed vertex-level embedding method, and generalizes it to obtain graph-level representations. To do so, WSGCN introduces additional \textit{master nodes} in the original network, that are connected to all the other vertices (akin to the virtual nodes used by Graphormer). When using a GCN to learn vertex-level representations, the vectors obtained for these specific global vertices can be used to represent the whole graph. WSGCN proposes five variants based on different connection schemes for the master nodes: \texttt{WSGCN+} (only positive edges), \texttt{WSGCN-} (only negative edges), \texttt{WSGCN±} (negative vs. positive master nodes), \texttt{WSGCNsb} (based on Structural Balance) and \texttt{WSGCNgb} (Structural Balance generalization).

%%%%%%%%%%%%%%%%%%%%%%%%%%%%%%%%%%%%%%%%%%%%%
\section{Proposed Graph Embedding Methods}
\label{sec:ProposedMethods}
Previous work~\cite{Cecillon2024phd} showed that integrating additional information into the conversational graph has a positive impact on the performance of abusive message detection. \textcolor{highlightCol}{Consequently, we want to take into account edge weights, signs, directions, and vertex attributes directly when training whole graph embedding methods, to enrich their learned representations. However, none of the approaches from the literature, and consequently none of the methods selected in Section~\ref{sec:EmbGraphMethods}, can deal with all these types of information.} 

To fill this methodological gap, in this section, we extend SG2V and WSGCN, two whole-graph embedding methods (cf. Section~\ref{sec:EmbGraphMethodsGr}), to take advantage of all the available information. We first introduce the vertex attributes used by our methods in Section~\ref{sec:NodeAttributes}. Then we present a Weighted Directed Attributed variant of SG2V (WDA-SG2V) in Section~\ref{sec:WDA-SG2V}, that generalizes the Weisfeiler--Lehman relabeling procedure to handle these supplementary edge attributes. Finally, we propose a Weighted Directed Attributed variant of WSGCN (WDA-WSGCN) in Section~\ref{sec:WDA-WSGCN}.

%%%%%%%%%%%%%%%%%%%
\subsection{Vertex attributes}
\label{sec:NodeAttributes}
\begin{figure}[htb!]
    \centering
    \resizebox{0.30\linewidth}{!}{\input{figures/TunedG2Vauthor}}
    \hfill
    \resizebox{0.30\linewidth}{!}{\input{figures/TunedG2Vdistance}}
    \hfill
    \resizebox{0.30\linewidth}{!}{\input{figures/TunedG2Vtarget}}
    \caption{Examples of the three vertex attributes: (a) \textit{author}; (b) \textit{distance}; and (c) \textit{target}. The targeted vertex is represented in red.}
    \label{fig:G2Vrelabeling}
\end{figure}
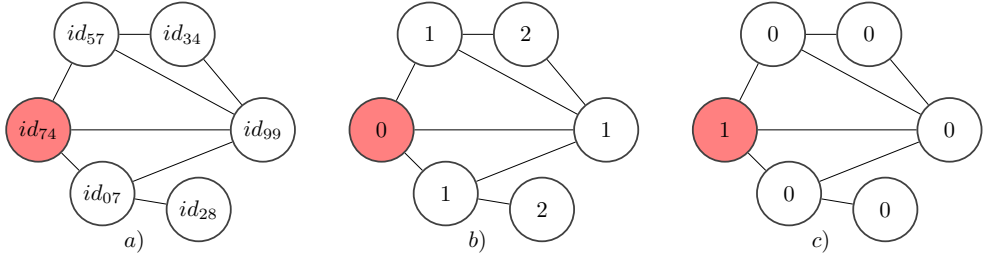

Vertex attributes allow the introduction of some user-specific information into the learned representation of the conversational graphs. We propose three different attributes that are specifically designed for our task and illustrated in Figure~\ref{fig:G2Vrelabeling}.

The first one, called \textit{author} (Figure~\ref{fig:G2Vrelabeling}.a), consists in using the unique identifier associated with the author of a post in our dataset. Therefore, we can identify if a same author is active in several conversations. Our intuition is that identifying the authors could allow us to consider context at a larger scale than a simple conversation. For instance, the fact that two authors interact in multiple conversations and that there is a history between them could be crucial to detecting abusive situations.

With \textit{distance} (Figure~\ref{fig:G2Vrelabeling}.b), we compute the geodesic distance between the considered vertex and the targeted vertex (which represents the author of the targeted message) and use it as its label. The purpose is to group different vertices (i.e. users) with the same label depending on their proximity with the author represented by the targeted vertex. The groups with low distances could correspond to the persons close to this author, while those with high distances represent the users that only have brief or indirect contact with him. This strategy reduces the number of distinct labels in the graph and could make it easier to detect structural equivalence.

Finally, \textit{target} (Figure~\ref{fig:G2Vrelabeling}.c) is an extension of the previous strategy that further reduces the number of labels down to two. This binary label indicates whether the vertex is the targeted one or not. Our intuition is that the targeted vertex plays a particular role in the graph, thus it could be important to differentiate it from others.

%%%%%%%%%%%%%%%%%%%
\subsection{Weighted Directed Attributed SG2V (WDA-SG2V)}
\label{sec:WDA-SG2V}
As explained in Section~\ref{sec:EmbGraphMethodsGr}, the original Graph2vec method requires a label to identify the vertices and apply the Weisfeiler--Lehman relabeling procedure. The default approach is to use the vertex degree as the default label. However, if the vertices can be described through some attribute, it is possible to use it as the label. This introduces another dimension of similarity between vertices, in addition to the graph structure. In our case, we use each one of the three attributes described in Section~\ref{sec:NodeAttributes} separately as a label, and we additionally consider all possible combinations of these attributes by concatenating them to produce composite labels.

SG2V~\cite{Cecillon2024} proposes two adaptations of the Graph2vec model relying on variants of the Weisfeiler--Lehman relabeling procedure able to handle edge signs. To incorporate supplementary edge attributes in the learned representations, we follow the same procedure and propose a new variant able to handle edge weights and directions. The original Graph2vec label update rule is an iterative process where a vertex is described by a tuple constituted of its previous label, and a sorted multiset containing those of its neighbors. To incorporate both other edge properties, we first consider the set of \textit{outgoing} neighbors of a vertex $u$ and split them into quartiles depending on the \textit{weight} of the concerned edges. Based on this breakdown, we can attribute a quartile number (i.e. 1, 2, 3, or 4) to each edge. This allows distinguishing vertices holding the same label but connected to $u$ with different weight levels. We experimentally determined that splitting the weights into quartiles is the best option, at least for our abuse detection task. Using fewer quantiles does not provide enough possibilities to distinguish edges, and using more leads to too many possible labels, hence reducing the efficiency of the method.

The first variant proposed in~\cite{Cecillon2024}, SG2Vn, uses the original label update rule except that it appends the sign of the concerned edge in front of each neighbor. Our proposition for \textit{WDA-SG2Vn} is to further append the quartile number in front of each neighbor when building the labels in the following manner:
\begin{equation}
    \ell_t(u) = f \Big( \ell_{t-1}(u), \Big\{ \big[q(u,v), s(u,v), \ell_{t-1}(v) \big] : v \in N(u) \Big\} \Big),
\end{equation}
where $q(u,v)$ and $s(u,v)$ are the quartile number and the sign of edge $(u,v)$, respectively, $[~]$ denotes string concatenation, $\ell_t(u)$ is the label of the subgraph rooted in $u$ at iteration $t$, $N(u)$ is the neighborhood of $u$, and $f$ is an injective function used to replace the tuples by new labels.

The second SG2V variant proposed in~\cite{Cecillon2024}, SG2Vsb, creates two labels for each vertex, based on its positive and negative reachable sets. Our proposition for \textit{WDA-SG2Vsb} is the same as above, i.e. adding the quartile number in front of each positive and negative neighbor when building the labels. The update rules for the positive and negative labels are:
\begin{align}
    \ell_t^+(u) &= f\Big(\ell_{t-1}^+(u), \Big\{ \big[q(u,v), \ell_{t-1}^+(v) \big] : v \in N^+(u) \Big\}, \Big\{ \big[q(u,v), \ell_{t-1}^-(v) \big] : v \in N^-(u) \Big\} \Big) \\
    \ell_t^-(u) &= f\Big(\ell_{t-1}^-(u), \Big\{ \big[q(u,v), \ell_{t-1}^-(v) \big] : v \in N^+(u) \Big\}, \Big\{ \big[q(u,v), \ell_{t-1}^+(v) \big] : v \in N^-(u) \Big\} \Big).
\end{align}

Once all the iterations are completed, $f$ is applied to tuples formed by the positive and negative labels of each vertex, resulting in the final rooted subgraph labels.

%%%%%%%%%%%%%%%%%%%
\subsection{Weighted Directed Attributed WSGCN (WDA-WSGCN)}
\label{sec:WDA-WSGCN}
The original SGCN model takes an input feature matrix $\textbf{X}$ which we directly use to introduce vertex attributes in the representation. For each vertex, we use the three attributes described in Section~\ref{sec:NodeAttributes}. In~\cite{Cecillon2024}, SGCN was adapted to handle whole graphs, resulting in WSGCN, by adding master nodes to the GCN and using their representations as the representation of the whole graph.

To make this model able to use edge weights and directions, we propose a modification of its aggregation step. The original model consists in collecting features from neighboring vertices and aggregating them through average. In WDA-WSGCN, to obtain the representation of each vertex $u$, we first consider only their set of \textit{ingoing} neighbors, as they are the ones that influence the vertex. Then, to integrate the fact that all these neighbors have different degrees of importance depending on the weight of the edge shared with the vertex of interest, we perform a weighted mean aggregation of the neighbors. In other words, we weigh the features from neighboring vertices by the edge weight that we normalize to obtain a sum of 1. Then we aggregate them. This gives more importance to neighbors with a strong relationship in the resulting representation. We apply this procedure with all five interconnection schemes proposed for WSGCN in~\cite{Cecillon2024} (\texttt{WSGCN+}, \texttt{WSGCN-}, \texttt{WSGCN±}, \texttt{WSGCNsb} and \texttt{WSGCNgb}).

%%%%%%%%%%%%%%%%%%%
\subsection{Overview of the Graph Embedding Methods}
\label{sec:EmbComp}
Table~\ref{tab:GraphEmbeddingProperties} lists the preexisting graph embedding methods described in Section~\ref{sec:EmbGraphMethods}, as well as the variants WDA-SG2V and WDA-WSGCN that we propose in Section~\ref{sec:ProposedMethods}. For each method, it indicates which information they handle, in addition to the bare graph structure. This includes edge-related information (weights, directions, and signs) and vertex attributes. Only our proposed methods can deal with all four types of information at once. 

\begin{table}[htb!]
    \centering
    \begin{tabular}{l l l c c c c}
        \toprule
        \textbf{Scale} & \textbf{Method} & \textbf{Reference} & \textbf{Weight} & \textbf{Direction} & \textbf{Sign} & \textbf{Attribute} \\
        \midrule
        Vertices & DeepWalk & \cite{Perozzi2014} & - & - & - & - \\
        & Node2vec & \cite{Grover2016} & \checkmark & \checkmark & - & -\\
        & Walklets & \cite{Perozzi2017} & - & - & - & -\\
        & BoostNE & \cite{Li2019} & \checkmark & - & - & -\\
        & GraphWave & \cite{Donnat2018} & - & - & - & - \\
        & KH-GNN & \cite{Nikolentzos2020} & - & - & - & - \\
        \midrule
        Whole graph & FGSD & \cite{Verma2017} & \checkmark & - & - & - \\
        & Spectral Features & \cite{deLara2018} & - & - & - & -\\
        & NGNN & \cite{Zhang2021} & - & - & - & - \\
        & Graphormer & \cite{Ying2021} & \checkmark & \checkmark & - & - \\
        & SG2V & \cite{Cecillon2024} & - & - & \checkmark & \checkmark \\
        & WDA-SG2V & Section~\ref{sec:WDA-SG2V} & \checkmark & \checkmark & \checkmark & \checkmark \\
        & WSGCN & \cite{Cecillon2024} & - & - & \checkmark & - \\
        & WDA-WSGCN & Section~\ref{sec:WDA-WSGCN} & \checkmark & \checkmark & \checkmark & \checkmark\\
        \botrule
    \end{tabular}
    \caption{Summary of the edge (weight, direction, and sign) and vertex (attribute) additional information handled by the graph embedding methods described in Sections~\ref{sec:EmbGraphMethods} and~\ref{sec:ProposedMethods}.}
    \label{tab:GraphEmbeddingProperties}
\end{table}

%%%%%%%%%%%%%%%%%%%%%%%%%%%%%%%%%%%%%%%%%%%%%
\section{Experimental Setup}
\label{sec:ExperimentalSetup}
In this section, we present our experimental setup to assess the selected and proposed embedding methods for the abuse detection task. We first describe our dataset in Section~\ref{subsec:Dataset}, before providing an overview of the processing steps constituting our experiments in Section~\ref{subsec:Overview}, and finally turning to the selected libraries and their parameters in Section~\ref{subsec:ExperimentalProtocol}.

%%%%%%%%%%%%%%%%%%%
\subsection{Dataset Description}
\label{subsec:Dataset}
In our experiments, we use a proprietary dataset called SpaceOrigin~\cite{Papegnies2017a}. It is a collection of annotated messages written in French, and enriched with their conversational context, i.e. the messages posted before and after in the same conversation. The messages were extracted from a database of in-game interactions between users of the massively multiplayer online role-playing game \textit{SpaceOrigin}\footnote{\href{https://www.spaceorigin.fr}{\texttt{https://www.spaceorigin.fr}}}. They were posted in the in-game chat used by all users to communicate, propose alliances, or make strategies. The \textit{Abuse} class is constituted of 655 messages which have been flagged as being abusive by at least one user, and confirmed as such by a human moderator. This ensures a high-quality annotation. Among the messages never flagged by a confirmed abuse report, 1,890 were chosen at random to compose the \textit{Non Abuse} class. Each of the 2,545 messages in the constituted dataset, whatever its class, is associated with its surrounding context (i.e. messages posted in the same thread). It is composed of up to 2,000 messages on each side of the message (before and after). The SpaceOrigin dataset thus provides 2,545 distinct conversations, each containing one annotated message and the context before and after it. The top part of Table~\ref{tab:DatasetStats} provides some statistics regarding these messages and conversations.

For our experiments with graph embedding approaches, we use networks that model these conversations. The bottom part of Table~\ref{tab:DatasetStats} provides some statistics regarding this part of our dataset. We extract these graphs using the method proposed in~\cite{Cecillon2024}, itself based on~\cite{Papegnies2019}. The original method produces directed weighted graphs. Each network is built around a message of interest, called the \textit{targeted message}, and aims at modeling its conversational context. The targeted message is the object of the classification. In these networks, vertices represent users, and weighted edges the intensity of their message exchanges. Each network integrates the messages present in a so-called \textit{context period}, which contains a fixed number of messages occurring right before and after the targeted message. Temporal integration is performed by sliding a fixed-sized window over the context period, and incrementing edge weights based on the co-occurrence of speakers in this window. To further extract edge signs, we extended the procedure to include a sentiment analyzer that infers the polarity of the interactions between users based on the content of the exchanged messages~\cite{Cecillon2024}. This polarity corresponds to the sign of the edge. \textcolor{highlightCol}{It is worth stressing that besides this last sentiment analysis step, the graph extraction process does not use the textual content of the exchanged messages. As a consequence, if this optional last step is omitted, our graph-based abuse detection method is language-independent.} 

\begin{table}[htb!]
    \color{highlightCol}
    \centering
    \begin{tabular}{l r@{~}l r r}
        \toprule
        & \multicolumn{2}{c}{\textbf{Average}} & \textbf{Minimum} & \textbf{Maximum} \\
        \midrule
        \textbf{Number of words by message} & 27.34 & ±7.88 & 1 & 199\\
        \textbf{Number of characters by message} & 159.36 & ±40.21 & 2 & 645\\
        \textbf{Number of messages by conversation} & 553.68 & ±126.91 & 12 & 2,400 \\
        \midrule
        \textbf{Number of vertices by graph} & 47.74 & ±20.34 & 2 & 214\\
        \textbf{Graph density} & 0.48 & ±0.16 & 0.10 & 1.00 \\
        \textbf{Number of negative edges by graph} & 166.1 & ±22.96 & 1 & 1,692 \\
        \textbf{Number of positive edges by graph} & 245.9 & ±30.41 & 1 & 2,323 \\
        \botrule
    \end{tabular}
    \caption{Statistics describing the messages (top part) and graphs (bottom part) of our dataset. Symbol $\pm$ denotes the standard deviation.}
    \label{tab:DatasetStats}
\end{table}

The conversation-oriented nature of our dataset and methods makes it difficult to compare our approach to existing abuse detection tools from the literature. On the one hand, according to a recent review~\cite{Cecillon2024phd}, all other available datasets annotated for abuse detection are constituted of independent messages, or very short sequences (4--5 messages) in a very few cases. Our dataset is the only one that provides long, full conversations annotated for abuse. Because our methods are designed to take advantage of such conversations, they cannot be applied to traditional benchmarks from the literature, which do not provide the required conversational context. 
On the other hand, and as a consequence of other datasets being message-oriented, all existing abuse detection methods are tailored to process independent messages. Applying them to our data would result in them ignoring most of the conversation, and would therefore be an unfair comparison. In addition, all these tools are designed and/or trained for the English language. According to our preliminary experiments, applying them to our French dataset results in a serious drop in performance, which is why we do not present any such results in this article. 

Another possible issue regarding our corpus is its relatively small size. We have conducted additional experiments in order to study the effect that the quantity of data used for training has on abuse detection. As expected, they showed that using more training data improves classification performance. However, these experiments also revealed that all the methods considered in this article are similarly affected by corpus size. This suggests that the observations made in Section~\ref{sec:ClassificationResults} would hold for larger datasets. 

For legal reasons related to the commercial nature of the raw textual data, we cannot share publicly the content of the exchanged messages. However, all the extracted networks are available online\footnote{\href{https://doi.org/10.5281/zenodo.11617245}{\texttt{https://doi.org/10.5281/zenodo.11617245}}}.

%%%%%%%%%%%%%%%%%%%
\subsection{\textcolor{highlightCol}{Processing Steps}}
\label{subsec:Overview}
\textcolor{highlightCol}{Our experiments aim at assessing the discriminative power of the different representation methods listed in Sections~\ref{sec:EmbeddingMethods} and~\ref{sec:ProposedMethods}, regarding the classification of abusive vs. non-abusive messages. We consider using them separately and combining them.} 

\begin{figure}[htb!]
    \color{highlightCol}
    \center
    \includegraphics[width=1\textwidth]{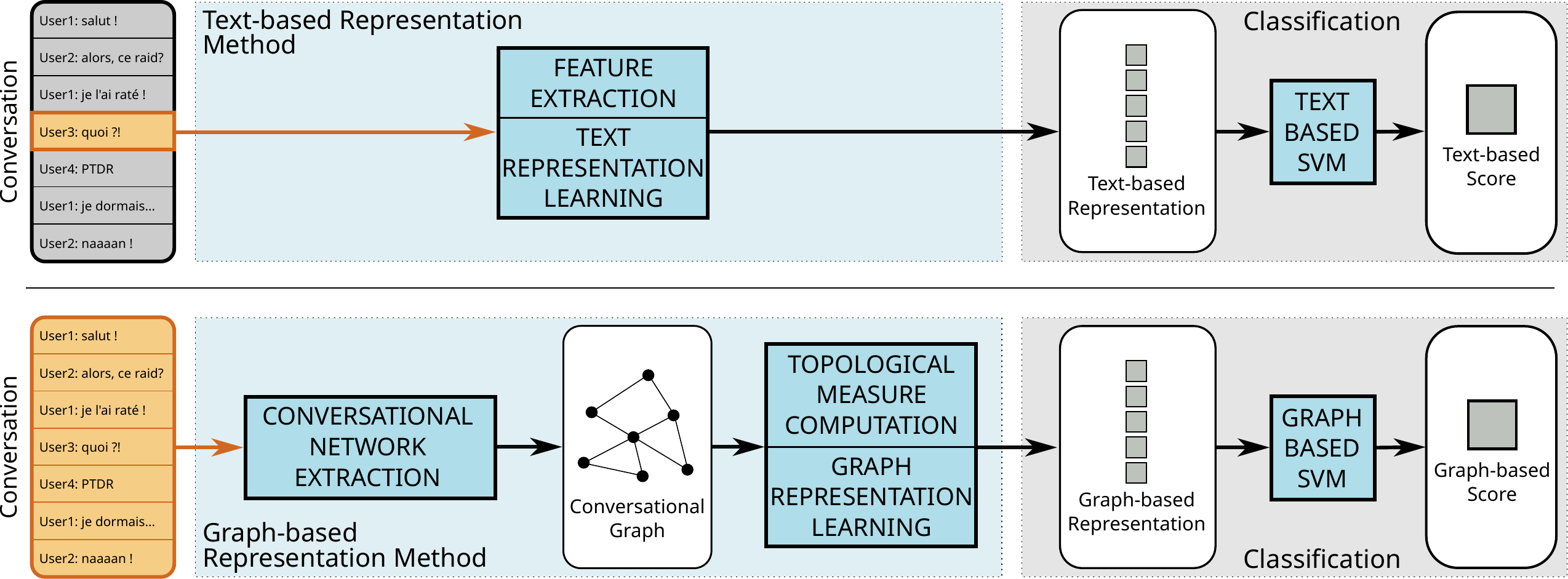}
    \caption{Processing steps performed for each representation method considered in the experiments. The top part focuses on text-based methods, which use only the classified message as input, whereas the bottom part focuses on graph-based methods, which use the whole conversation, and include an additional graph extraction step. In both cases, the baseline relies on feature engineering, whereas all the other methods considered here use representation learning. The SVM-based classification phase is the same for text- and graph-based methods.}
    \label{fig:MainPipeline}
\end{figure}

%%%
\medskip\noindent\textcolor{highlightCol}{\textbf{Separate Representations}}
\textcolor{highlightCol}{Figure~\ref{fig:MainPipeline} shows the different processing steps used to leverage these methods and perform the abuse detection task. It distinguishes between text-based methods (top part of the figure) and graph-based ones (bottom). The former focus only on the targeted message, which we want to classify. The selected representation learning methods are directly applied to the text, using the implementations and parameters later specified in Section~\ref{subsec:ExperimentalProtocol}. As a baseline, we also include the feature-based method developed in~\cite{Papegnies2017}. It uses appropriate measures selected through a \textit{Feature engineering} process, including the message and word lengths, the classes and number of unique characters, the number of capital letters, the compression ratio of the message, the number of repeated characters, the number of words and bad words, two TF-IDF scores and a probability score based on a bag-of-words representation of the messages.} 

\textcolor{highlightCol}{By comparison, the graph-based methods consider the whole conversation, and they require an additional step in order to extract the conversational graph as explained in Section~\ref{subsec:Dataset}. The selected graph representation methods are then directly applied to the graphs, using the libraries and parameters indicated in Section~\ref{subsec:ExperimentalProtocol}. Like for text, we also compute the features developed in~\cite{Papegnies2019, Cecillon2021} through a feature engineering process, as a baseline. They include hundreds of topological measures, defined at different scales and leveraging different scopes to represent the graphs. Most are standard complex network measures: standard degree, betweenness, closeness, modularity, etc. For the sake of space, we do not list them in this paper and refer the interested reader to~\cite{Papegnies2019, Cecillon2024phd} for a formal definition of the measures.} 

\textcolor{highlightCol}{In all cases (text- and graph-based methods, feature engineering and representation learning), the obtained message or conversation representation is fetched to an SVM classifier, which output a score used to determine whether the targeted message is abusive or not.} 

%%%
\medskip\noindent\textcolor{highlightCol}{\textbf{Representation Combination}}
\textcolor{highlightCol}{As already mentioned before, in this work we also want to study the impact that combining pairs of representations methods has on abuse detection. Previous studies in the literature~\cite{Mishra2018b,Cecillon2019} have shown that combining multiple types of information can be effective in the context of abusive message detection. For this reason, we propose to combine the most competitive text and graph embedding methods. Our intuition is that these methods, based on completely different modalities, namely the text and the conversational graphs, capture different, possibly complementary, information. Therefore, their combination should improve classification performance. Additionally, we consider combining methods that operate on the same modality. As mentioned by \citeauthor{Le2020en}~\cite{Le2020en}, the combination of multiple methods can lead to enhanced performance.} 

\begin{figure}[htb!]
    \color{highlightCol}
    \center
    \includegraphics[width=1\textwidth]{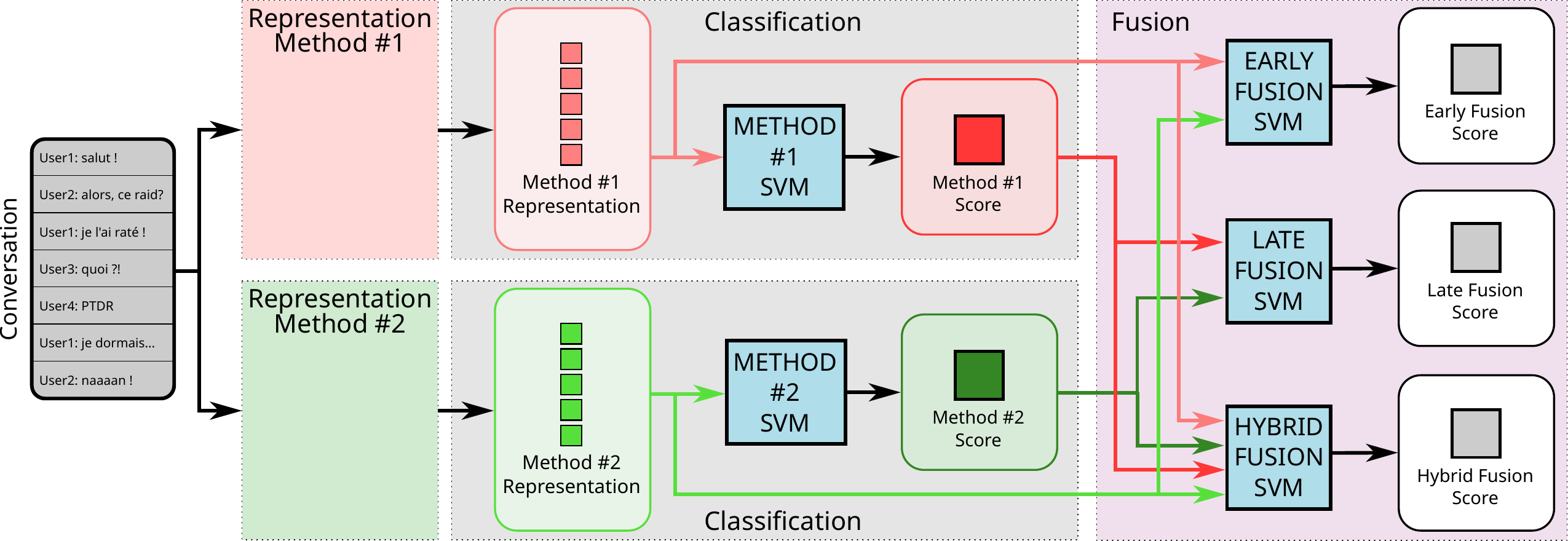}
    \caption{Illustration of the three fusion strategies used to combine pairs of representations. On the left, the red and green blocks correspond to two representation methods of interest, each one outputting some vector representation fetched to an SVM classifier, similarly to what is shown in Figure~\ref{fig:MainPipeline}. The new part is the Fusion phase, displayed on the right, which involves three SVM using different inputs (see text).}
    \label{fig:FusionPipeline}
\end{figure}

\textcolor{highlightCol}{To combine these methods, we use the three fusion strategies proposed in~\cite{Cecillon2019}. These are represented in Figure~\ref{fig:FusionPipeline} for two generic representation methods, one shown in red, the other in green. These strategies rely on the representations produced when considering these methods separately, as well as on the classifiers already trained using these representations, which are denoted by \textit{Method \#1 SVM} and \textit{Method \#2 SVM} in the figure, respectively. First, \textit{Early Fusion} involves concatenating the representations produced by both methods in order to train a classifier, denoted by \textit{Early Fusion SVM} in the figure. Second, \textit{Late Fusion} consists in leveraging the outputs of the classifiers trained independently on the two representations (dark red and green, in the figure), in order to use them as inputs when training a third classifier, denoted by \textit{Late Fusion SVM}. Finally, \textit{Hybrid Fusion} simultaneously uses all the inputs of both other strategies to train a new classifier, i.e. the representations as well as the output scores of the original SVMs.}

%%%%%%%%%%%%%%%%%%%
\subsection{Software and Parameters}
\label{subsec:ExperimentalProtocol}
We first review the operational settings used with the text and graph embedding methods previously described, before turning to the classification step that takes advantage of the representations they produce to distinguish between abusive and non-abusive messages. Our source code is available online\footnote{\href{https://github.com/CompNet/AlertEmbeddings}{\texttt{https://github.com/CompNet/AlertEmbeddings}}}, \textcolor{highlightCol}{as well as the main models trained during our experiments. In the following, we focus only on the most important hyperparameters, but the comprehensive list is available on the same repository.} We conduct our experiments on an Nvidia RTX 2080 Ti GPU. 

%%%
\medskip\noindent\textbf{Text Embeddings} 
The textual aspect relies on the content of the messages to perform the detection. We use the five text embedding models described previously in Section~\ref{sec:EmbLexicalMethods}. The five approaches generate embeddings at the word level. As our objective is to represent the message as a whole and not individual words, we use the standard technique of averaging the representations of all words in a message to obtain its global embedding~\cite{Ruckle2018}. Furthermore, the methods based on the BERT architecture, i.e. CamemBERT and FlauBERT, directly learn a global representation of the sequence through a particular token added at the very beginning of the sequence. We use this global representation as a second technique to represent the messages, in addition to the averaged sequence representations. 

We use models trained on French text. We keep the original dimension of the vector representations for all the pre-trained models, which are listed in Table~\ref{tab:EmbeddingPerfLexical}. The Word2vec\footnote{\href{https://github.com/Kyubyong/wordvectors}{\texttt{https://github.com/Kyubyong/wordvectors}}}, fastText\footnote{\href{https://github.com/flairNLP/flair/blob/master/resources/docs/embeddings/CLASSIC_WORD_EMBEDDINGS.md}{\texttt{https://github.com/flairNLP/fasttext}}} and Flair\footnote{\href{https://github.com/flairNLP/flair/blob/master/resources/docs/embeddings/FLAIR_EMBEDDINGS.md}{\texttt{https://github.com/flairNLP/flair}}} models all use text extracted from Wikipedia.  

CamemBERT's authors follow the original architecture configurations of BERT. They propose a CamemBERT$_{\text{BASE}}$ model with 110M parameters, and a CamemBERT$_{\text{LARGE}}$ model with 24 layers of encoders, 1,024 hidden dimensions, 16 attention heads, and 335M parameters. According to the authors' study~\cite{Martin2020en}, the latter gives better results on named entity recognition but similar results on part-of-speech and dependency parsing tasks. Thus, we use the CamemBERT$_{\text{LARGE}}$ model in the remainder of this article. This model\footnote{\href{https://camembert-model.fr/}{\texttt{https://camembert-model.fr/}}} is pre-trained on the \textit{CCNet}~\cite{Wenzek2020} corpus containing 135 GB of French text extracted from various websites.

FlauBERT~\cite{Le2020en} also proposes a FlauBERT$_{\text{BASE}}$ model and a FlauBERT$_{\text{LARGE}}$ model with 24 layers of encoders, 1,024 hidden dimensions, and 16 attention heads. These models are trained on a corpus of 71 GB of French text aggregated from multiple online sources. We use the FlauBERT$_{\text{LARGE}}$ model\footnote{\href{https://github.com/getalp/Flaubert}{\texttt{https://github.com/getalp/Flaubert}}}.

%%%
\medskip\noindent\textbf{Graph Embeddings}
The structural aspect focuses on the use of contextual information. It relies on the conversational graphs extracted from the SpaceOrigin dataset. They represent the full conversations (i.e. context) in which messages were posted. In the following experiments, we use the existing embedding methods described in Section~\ref{sec:EmbGraphMethods} and the two proposed in~\ref{sec:ProposedMethods}. For the existing methods, all implementations we use come from the \textit{Karate Club Toolkit}~\cite{Rozemberczki2020}, except for Node2vec, for which we use the implementation released by E. Cohen\footnote{\href{https://github.com/eliorc/node2vec}{\texttt{https://github.com/eliorc/node2vec}}}, and Graphormer, for which we use the original implementation released by its authors\footnote{\href{https://github.com/microsoft/Graphormer}{\texttt{https://github.com/microsoft/Graphormer}}}. The dimensions of all the vector representations are listed in Table~\ref{tab:EmbeddingPerfGraph}.

Half of the selected methods learn \textit{vertex} embeddings: we use only the representation of the vertex corresponding to the author of the message that we want to classify. We also experiment by averaging the representations over all the graph vertices to get a single global representation, however, it is much less effective, so we do not report these results here. The other half of the selected methods directly provide a representation of the whole graph.

%%%
\medskip\noindent\textbf{Classification}
For the classification step, we use an SVM implemented in the Sklearn toolkit~\cite{Pedregosa2011} under the name SVC (C-Support Vector Classification). This classifier is well-suited for small datasets such as ours. 

We set up our experiments with a 10-fold cross-validation. We use a 70~\%-train / 30~\%-test split, which means that for each run of the cross-validation, 7 folds are used for training and the remaining 3 compose the test set. All the results are expressed in terms of macro $F$-measure, the unweighted arithmetic mean of all the per-class $F$-measures. This metric allows handling all classes equally, even with our unbalanced dataset. \textcolor{highlightCol}{The $F$-measure is itself the harmonic mean of Precision and Recall, which we provide separately in Appendix~\ref{sec:Appendix}, for the sake of completeness.}

%%%%%%%%%%%%%%%%%%%%%%%%%%%%%%%%%%%%%%%%%%%%%
\section{Classification Results}
\label{sec:ClassificationResults}
In this section, we present and discuss the classification results and performance obtained with the methods described previously. We first consider text and graph embedding methods separately (Section~\ref{subsec:ExperimentsResults}), before combining them to assess their complementarity (Section~\ref{subsec:EmbeddingFusion}).

%%%%%%%%%%%%%%%%%%%
\subsection{Separate Modalities}
\label{subsec:ExperimentsResults}
We consider separately our experiments with text and graph embeddings on the abuse detection task. We also include the content- and graph-based original methods developed through feature engineering~\cite{Cecillon2024phd}, as baselines.

%%%%%%%%%
\subsubsection{Text Embeddings}
\label{subsec:ExperimentsResultsText}
Table~\ref{tab:EmbeddingPerfLexical} shows the results obtained by the baseline and the five selected text embedding methods. The first column indicates the scale of the generated representations: \textit{Word} for the methods that generate representations of words that we average to obtain the representation of a message, and \textit{Message} for BERT-based methods able to directly learn a global representation of the message. The third column shows the dimension of the learned representations. 

In many models, the predominant classification error involves incorrectly labeling \textit{non-abusive} messages as \textit{abusive}. This phenomenon represents approximately 60~\% of the total classification errors. Some non-abusive messages are quite ambiguous and therefore, difficult to classify.

\begin{table}[htb!]
    \centering
    \begin{tabular}{l l r r r }
        \toprule
        \textbf{Scale} & \textbf{Method} & \textbf{Dimension} & \textbf{$F$-measure} & \textbf{Total Runtime} \\
        \midrule
            -- & Baseline-Text & 29 & 75.21 & 0:41\\
        \midrule
            Word & Word2vec & 300 & 71.14 & 4:14\\
            & fastText & 300 & 76.01 & 4:29\\
            & Flair & 2,048 & 77.33 & 5:25 \\
            & CamemBERT-W & 1,024 & \textbf{81.02} & 6:56\\
            & FlauBERT-W & 1,024 & 78.46 & 7:11\\
        \midrule
            Message & CamemBERT-M & 1,024 & 80.96 & 4:59\\
            & FlauBERT-M & 1,024 & 74.99 & 5:08\\ 
        \botrule
    \end{tabular}
    \caption{$F$-measures obtained by the baseline approach (based on feature engineering) and the five word embedding methods described in Section~\ref{sec:EmbLexicalMethods}. The runtime is expressed as \textit{minutes:seconds}.  The runtime is expressed as hour:minute:seconds. \textcolor{highlightCol}{Table~\ref{tab:EmbeddingPerfLexicalPrecRec} in Appendix~\ref{sec:Appendix} shows the corresponding Precision and Recall scores.}}
    \label{tab:EmbeddingPerfLexical}
\end{table}

Our first observation is that all but one word embedding approach achieves a better $F$-measure than the baseline. Word2vec and fastText are among the least effective approaches for this classification task. This is not surprising, as they are the only methods that produce \textit{fixed} representations of words, i.e. a single representation for each word, regardless of context. Consequently, they are unable to distinguish homographs, which can lead to confusion regarding the meaning of a message. The performance gap between these two methods can be attributed to the fact that Word2vec is entirely incapable of handling out-of-vocabulary words, whereas fastText, by design, can manage them. In the context of online messages, this capability is crucial, given the prevalence of spelling mistakes, rare and highly specific terms, and even intentional obfuscation.

The contextualized embeddings learned by Flair yield better performance than fixed representations, but they are inferior to all other contextual methods. It appears that Flair's character-level representation is not well-suited for the abuse detection task, or at least not for our dataset.

FlauBERT-M, which learns a global representation of the message, gets lower performances than its counterpart Flaubert-W, which creates the embedding by averaging the representations of words composing the message. We assume that the abusive nature of a message is often based on just a few words, and that operating directly on words allows improving their detection. However, we do not find this situation with CamemBERT, as both its variants obtain performances that are not statistically different. CamemBERT-W, nonetheless, gets the best $F$-measure (81.02~\%) of all the text embedding methods. This constitutes an improvement of almost 6 points compared to our baseline.

These results confirm the potential of word embedding methods. On this classification task, they can learn representations that capture much more discriminative information compared to the baseline that relies on a handcrafted set of morphological and language features computed on the message. This directly leads to an improvement of up to 6 points in terms of $F$-measure on the abuse detection task. Furthermore, these methods are not built following the same concepts and objectives, therefore, we can suppose that they could be complementary. For instance, \citeauthor{Le2020en}~\cite{Le2020en} show that the combination of CamemBERT and FlauBERT yields better performances than the two methods used separately on a POS tagging task. We explore these options in Section~\ref{subsec:EmbeddingFusion}.

On the downside, word embedding methods require much more time than the baseline to obtain the representations of messages. This could be a limitation in a real-world application where the time to classify a message is limited.

%%%%%%%%%
\subsubsection{Graph Embeddings}
\label{subsec:ExperimentsResultsGraph}
In this section, we first discuss the results of the different graph embedding methods. We then focus on the impact of vertex and edge attributes, before discussing the effect of the different representation scales considered. 
The results of the graph embedding methods and the baseline are shown in Table~\ref{tab:EmbeddingPerfGraph}. The first column indicates whether it is a \textit{vertex} or \textit{whole-graph} embedding method. 

\begin{table}[htb!]
    \centering
    \begin{tabular}{l l r r r}
        \toprule
        \textbf{Scale} & \textbf{Method} & \textbf{Dimension} & \textbf{$F$-measure} & \textbf{Total Runtime}\\[2pt]
        \midrule
           -- & Baseline-Graph & 477 & 83.40 & 3:51:25 \\
        \midrule
            Targeted vertex & DeepWalk & 128 & 73.72 & 14:29 \\
            & Node2vec & 128 & 74.59 & 15:01 \\
            & Walklets & 128 & 74.84 & 16:21 \\
            & BoostNE & 136 & 71.53 & 16:00\\
            & GraphWave & 200 & 80.34 & 15:12 \\
            & $k$-hop GNN & 64 & 73.96 & 24:55 \\
        \midrule
            Whole graph & FGSD & 200 & 71.13 & 6:35 \\
            & Spectral Features & 128 & 74.23 & 6:57\\
            & NGNN & 64 & 74.51 & 16:58 \\
            & Graphormer & 768 & 73.87 & 14:11 \\
            & Graph2vec & 128 & 78.17 & 4:54 \\
            & SG2V & 128 & 79.13 & 6:24 \\ 
            & WDA-SG2V & 128 & \textbf{80.61} & 7:02\\ 
            & WSGCN & 128 & 74.74 & 14:14 \\
            & WDA-WSGCN & 128 & 75.80 & 15:07 \\ [2pt]
        \botrule
    \end{tabular}
    \caption{$F$-measures of the baseline and graph embedding methods, including the 13 selected in Section~\ref{sec:EmbGraphMethods}, and the two proposed in Section~\ref{sec:ProposedMethods}. The runtime is expressed as \textit{hour:minute:seconds}. \textcolor{highlightCol}{Table~\ref{tab:EmbeddingPerfGraphPrecRec} in Appendix~\ref{sec:Appendix} shows the corresponding Precision and Recall scores.}}
    \label{tab:EmbeddingPerfGraph}
\end{table}

%%%
\medskip\noindent\textbf{General Observations}
Unlike our observations with textual embeddings, no graph embedding method outperforms the baseline. This is not surprising, as the baseline feature set was specifically designed for this task and dataset, incorporating a very large number of topological measures that cover a wide range of graph-related concepts. In contrast, the embedding methods are relatively generic. Furthermore, the text baseline includes far fewer features than the graph baseline, which likely makes it easier to outperform. 

The Graph2vec variants WDA-SG2V and SG2V (both whole-graph scale approaches), GraphWave (vertex scale), and standard Graph2vec (whole-graph scale) yield the best performances among graph embedding approaches. \textcolor{highlightCol}{GraphWave is also the best \textit{language-independent} method: as explained in Section~\ref{subsec:Dataset}, our graph-based abuse detection approach is language-independent, provided one omits the sentiment analysis step of the graph extraction process, which allows assigning signs to the graph edges. GraphWave dominates all other methods that do not leverage edge signs.} 

The other methods obtain correct performances with an $F$-measure approximately ranging from 71~\% to 78~\%. Spectral Features and FGSD, which operate on the whole graph, might be penalized by the small size of our dataset and by the fact that graphs have approximately the same size and possibly similar structures. DeepWalk is less efficient than Node2vec, a result in line with other studies~\cite{Goyal2018, Grover2016}. The Walklets algorithm learns multi-scale relationships in the graph. However, such relationships might not be very developed in our graphs, which could explain its lower performance. This observation could also be the reason for the very poor performance of BoostNE, which also operates on several different granularity levels. 

In the end, the best graph embedding method is less efficient than the baseline by approximately 3 points. However, embedding-based approaches still have two major advantages compared to the feature-based baseline. First, they are not specifically designed for this task or dataset and are hence more likely to be efficient in other settings. For instance, the textual embedding models that we use are not specifically pre-trained on abusive datasets, which illustrates this ability to be applied in various contexts. Second, embedding methods are more scalable than hand-crafted sets of features. Computing the topological measures used in our baseline is computationally very expensive, with a total runtime of almost 4 hours. On the other hand, it only takes a few minutes, in the same conditions, to deal with the embedding methods on the same machine, which makes them a lot more time-efficient. 

%%%
\medskip\noindent\textbf{Impact of Edge and Vertex Attributes}
Our proposed method Weighted Directed Attributed Signed Graph2vec (WDA-SG2V) is two points above Graph2vec, which highlights the importance of using additional information for learning the representations. The experiments in~\cite{Papegnies2019} already showed that edge weights and directions allow improving the performance in feature-based methods: our results indicate that the same observation applies to embedding methods. Furthermore, edge signs and vertex attributes also help improve the learned representations. The proposed models are therefore able to effectively incorporate additional information to learn better representations of the graphs. This observation is confirmed by our second proposed method (WDA-WSGCN) which is one point above WSGCN.

This is an important result, as it illustrates the importance of including additional information in the representation learning process, at least for the task considered here. 

%%%
\medskip\noindent\textbf{Representation Scale}
We evaluate two different strategies in Table~\ref{tab:EmbeddingPerfGraph}: learning a representation of the entire graph and using the representation of the targeted vertex. For the latter, we assumed that the targeted vertex is the most important for our classification task.

Between vertex and whole-graph approaches, there is no clear distinction in terms of $F$-measure. Since the graph in its entirety represents the message and its associated conversation, one could assume that embedding the whole graph could allow capturing more information than a single vertex embedding. However, it seems that these graphs can be well-characterized by focusing on the vertex of the author of the \textit{targeted message}. This suggests that the relative position of this vertex in the graph provides enough information to characterize the entire conversation, and that vertex-level embeddings can effectively capture this information.

%%%%%%%%%
\subsubsection{\textcolor{highlightCol}{Qualitative Comparison}}
\label{subsec:ExperimentsResultsQuali}
\textcolor{highlightCol}{In order to understand what differentiates the text- and graph-based approaches in terms of classification behavior, we conduct a brief qualitative review of the cases where they disagree, inspired by~\cite{Dar2024}. We focus only on CamemBERT-W and WDA-SG2V, respectively the best text- and graph-based approaches according to our experiments, but our discussion applies more broadly to the other methods assessed in this article. A few particularly illustrative examples are described in Table~\ref{tab:QualiComp}, providing a glimpse as to which situations result in one approach bettering the other.} 

\begin{table}[htb!]
    \color{highlightCol}
    \centering
    \begin{tabular}{p{2.5cm} p{2.5cm} p{6.75cm}}
        \toprule
        \textbf{Original} & \textbf{English} & \textbf{Description} \\
        \textbf{Message} & \textbf{Translation} & \\
        \midrule
        \textit{AUCUN RESPECT WTF} ? & NO RESPECT WTF & The excessive use of upper case text is particularly revealing of the abuse, which explains why it could be better captured through the content-based approaches. \\[2mm]
        % \textit{bonjour les gros noobs} & Hello, you big noobs! & C'est assez explicite mais ça n'a pas généré de réactions chez les autres utilisateurs donc les graphes n'arrivent pas à le détecter. D'ailleurs ça ressemble à un message envoyé à des gens qu'il connaît donc sans agressivité et qui n'aurait peut-être pas dû être annoté comme abusif. \\
        \textit{Tu dégages de mon salon} & Get the hell out of my chat & This message was flagged \textit{a posteriori}: at the time of the conversation, it went completely unnoticed, and elicited no reactions from the other users, which simply continued their discussion. As a consequence, it had no effect on the conversation structure, and remained undetectable by graph-based approaches. \\
        \midrule
        \textit{dédicace à Yuuki !} & shoutout to Yuuki! & This message was not flagged due to its content, but because its author repeated it many times in succession. This form of spamming is undetectable without the conversational context, which is why only the graph-based methods are able to identify it. \\[2mm]
        \textit{laisse les grands parler} & let the grown-ups talk & There is no outright rudeness or vulgarity here, but rather a disdainful tone, which constitutes a form of implicit abuse that is difficult to detect from the text alone. The larger conversation reveals that two users are beginning a heated exchange, with tension building, which ultimately impacts the conversational graph. \\
        \botrule
    \end{tabular}
    \caption{Four examples of messages flagged as Abusive in the SpaceOrigin dataset. The first two are detected only by the text-based method (top half of the table), and the last two only by the graph-based method (bottom half).}
    \label{tab:QualiComp}
\end{table}

\textcolor{highlightCol}{Abusive messages detected by text-based methods but not by graph-based methods typically involve situations where the issue stems from a disregard for chatroom rules, yet has no visible effect on the conversation. The lack of reaction from other users may be due to them not perceiving the breach as true abuse, or from acknowledging it but intentionally ignoring the provocation. 
Abuse cases detected only by graph-based methods often involve more subtle situations, such as irony, sarcasm, or innuendo. These tend to be overlooked by text-based methods, likely due to their difficulty in capturing implied or secondary meaning. In contrast, while graph-based methods do not directly detect this kind of meaning either, one could argue that they do so indirectly, using the reactions of users as a proxy.}

%%%%%%%%%%%%%%%%%%%
\subsection{Fusion of Embeddings}
\label{subsec:EmbeddingFusion}
After the assessment of individual representation methods, we now turn to combinations of pairs of methods. We focus on the embedding methods that obtained the best performances when used separately in the previous sections. For the textual embeddings, we select Camembert-W, FlauBERT-W (word level), and CamemBERT-M (message level). For the graphs, we use GraphWave (vertex-scale approach) and WDA-SG2V (whole-graph scale approach). We also consider the baselines for both text- and graph-based methods. Table~\ref{tab:EmbeddingFusion} shows the $F$-measures obtained when applying the three fusion strategies to the best-performing methods.

\begin{table}[htb!]
    \centering
    \begin{tabular}{l l r r r r}
        \toprule
        \textbf{Fusion} & \textbf{Methods} & \textbf{Best} &\textbf{Dim.} & \textbf{$F$-measure} & \textcolor{highlightCol}{\textbf{Runtime}} \\
        \midrule
            Early & Graph Base + Text Base. & 83.40 & 506 & 84.51 & \textcolor{highlightCol}{3:55:17} \\
        \cline{2-6}
            & CamemBERT-W + CamemBERT-M & 81.02 & 2,048 & 81.32 & \textcolor{highlightCol}{14:37} \\
            & CamemBERT-W + FlauBERT-W & 81.02 & 2,048 & 79.85 & \textcolor{highlightCol}{16:12} \\
        \cline{2-6}
            & WDA-SG2V + GraphWave & 80.61 & 256 & 82.57 & \textcolor{highlightCol}{25:31} \\
        \cline{2-6}
            & WDA-SG2V + CamemBERT-W & 81.02 & 1,152 & 85.84 & \textcolor{highlightCol}{17:21} \\
            & GraphWave + CamemBERT-W & 81.02 & 1,152 & 85.99 & \textcolor{highlightCol}{24:59} \\
            & Graph Base. + CamemBERT-W & 83.40 & 1,501 & 86.77 & \textcolor{highlightCol}{4:01:26} \\ 
            & WDA-SG2V + Text Base & 80.61 & 157 & 82.39 & \textcolor{highlightCol}{9:47} \\ 
        \midrule
            Late & Graph Base. + Text Base. & 83.40 & 2 & 84.17 & \textcolor{highlightCol}{3:54:57} \\
        \cline{2-6}
            & CamemBERT-W + CamemBERT-M & 81.02 & 2 & 80.21 & \textcolor{highlightCol}{14:24} \\
            & CamemBERT-W + FlauBERT-W & 81.02 & 2 & 80.10 & \textcolor{highlightCol}{16:03} \\
        \cline{2-6}
            & WDA-SG2V + GraphWave & 80.61 & 2 & 82.19 & \textcolor{highlightCol}{25:32} \\
        \cline{2-6}
            & WDA-SG2V + CamemBERT-W & 81.02 & 2 & 85.29 & \textcolor{highlightCol}{16:29} \\
            & GraphWave + CamemBERT-W & 81.02 & 2 & 85.80 & \textcolor{highlightCol}{24:46} \\
            & Graph Base. + CamemBERT-W & 83.40 & 2 & 86.21 & \textcolor{highlightCol}{4:00:23} \\ 
            & WDA-SG2V + Text Base & 80.61 & 2 & 82.27 & \textcolor{highlightCol}{9:33} \\ 
        \midrule
            Hybrid & Graph Base. + Text Base. & 83.40 & 508 & 84.97 & \textcolor{highlightCol}{3:57:44} \\
        \cline{2-6}
            & CamemBERT-W + CamemBERT-M & 81.02 & 2,050 & 81.17 & \textcolor{highlightCol}{15:02} \\
            & CamemBERT-W + FlauBERT-W & 81.02 & 2,050 & 80.34 & \textcolor{highlightCol}{16:59} \\
        \cline{2-6}
            & WDA-SG2V + GraphWave & 80.61 & 258 & 81.34 & \textcolor{highlightCol}{27:44} \\
        \cline{2-6}
            & WDA-SG2V + CamemBERT-W & 81.02 & 1,154 & 86.27 & \textcolor{highlightCol}{19:17} \\
            & GraphWave + CamemBERT-W & 81.02 & 1,154 & 86.11 & \textcolor{highlightCol}{27:38} \\
            & Graph Base. + CamemBERT-W & 83.40 & 1,503 & \textbf{87.06} & \textcolor{highlightCol}{4:02:56}\\ 
            & WDA-SG2V + Text Base & 80.61 & 159 & 82.55 & \textcolor{highlightCol}{10:25} \\
        \botrule
    \end{tabular}
    \caption{$F$-measures obtained by the fusion of the best performing embedding methods following the three fusion strategies (Early, Late, Hybrid). Column \textit{Best} indicates the performance of the best method among the two combined ones, \textit{Dim.} denotes the dimension of the vector representation, \textit{$F$-measure} is the performance, and \textit{Runtime} is the total runtime (\textit{hour:minute:seconds}). The table is split horizontally into three parts, each one dedicated to one fusion strategy. In each such part, the same methods are gathered in four groups, depending on the nature of the combined methods. From top to bottom: baselines, fusion of text embeddings, fusion of graph embeddings, and fusion of text and graph embeddings. \textcolor{highlightCol}{Table~\ref{tab:EmbeddingFusionPrecRec} in Appendix~\ref{sec:Appendix} shows the corresponding Precision and Recall scores.}}
    \label{tab:EmbeddingFusion}
\end{table}

Our first observation is that all three fusion strategies yield fairly similar results. Generally speaking, hybrid fusion achieves top performance, ahead of early fusion and late fusion. This outcome is expected since hybrid fusion combines the other two methods. With the late fusion strategy, all the information available in early fusion is summarized into just two scores. One might assume that such compression would cause a serious loss of  discriminant information. Yet, despite this, late fusion performs only slightly worse than early fusion. Moreover, the systematic improvement obtained with hybrid fusion (compared to early fusion) shows that the two scores from late fusion can sometimes help the classifier. We hypothesize that, in certain cases, the relatively large representations used for early fusion contain some noise that affects classification performance. The late fusion scores can be seen as a more compact and less noisy representation, which is more suitable to these specific cases. 

The hybrid fusion of the two baselines improves the performance by more than 2 points compared to the graph baseline used alone. The fusion of the two CamemBERT approaches (CamemBERT-W and CamemBERT-M) does not bring a significant performance gain. Therefore, we suppose that they both capture similar information. The same applies to the fusion of CamemBERT and FlauBERT, which stays on par with the performance of CamemBERT used on its own. These text embeddings thus capture redundant information, which can be explained by the fact that they are both based on the same RoBERTa architecture. 

For the graph-based methods, the early fusion of GraphWave and WDA-SG2V also results in a clear improvement (82.57~\%), reaching a performance only one point lower than the graph baseline (83.40~\%). On the one hand, this result acknowledges the assumption that graph embedding methods that operate on different granularity levels (vertices for GraphWave, whole-graph for WDA-SG2V), can capture complementary information. On the other hand, the late and hybrid fusions perform worse. Thus, we hypothesize that all the complementary information captured by these two methods cannot be summarized using only two scores. 

Finally, the combination of the best text-based method (CamemBERT-W) with each approach based on graphs always achieves better performance than the combination of the two baseline methods (84.97~\%). The hybrid fusions of CamemBERT-W with WDA-SG2V and GraphWave obtain an $F$-measure of 86.27~\% and 86.11~\% respectively. The overall best performance is achieved by combining CamemBERT-W with the graph baseline (87.06~\%). Interestingly, this combination is only one point above both previous combinations, while when used on their own, WDA-SG2V and GraphWave are 5 points behind the baseline. The best combination method is 7 points above the best embedding method used alone. This systematic improvement of the performance obtained when combining text and graph information confirms previous results~\cite{Papegnies2019}. Our interpretation of this result is that, in certain cases, the textual content of the exchanged messages is enough to distinguish a normal from an abusive situation; whereas; in other cases, the text is not discriminant, but certain changes in the structure of the conversation are sufficient to discriminate. In most situations, text and structure are redundant, which is why the performance is good when using both types of information separately. In a minority (but significant proportion) of cases, though, they are not, and using both helps to improve classification performance.

%%%%%%%%%%%%%%%%%%%
\subsection{Results Summary}
\label{sec:EmbeddingResultsSummary}
To summarize our results presented in this section, Table~\ref{tab:EmbeddingsBestPerfs} shows the best performance we obtain for each type of approach studied. For text embeddings, the best method is CamemBERT applied to the words (CamemBERT-W) and averaged over the set of words composing a message. For graphs, it is WDA-SG2V, one of the whole-graph embedding methods that we propose. 
The results presented in Sections~\ref{subsec:ExperimentsResultsText} and~\ref{subsec:ExperimentsResultsGraph} allow answering RQ1 (How do feature engineering and representation learning methods compare in terms of abuse detection performance?). For the textual content, embedding methods clearly outperform the feature engineering approach. On the contrary, for the graph structure, the latter obtains better performance, but the gap with the best graph embedding methods is much smaller than for text. In both case, this performance gain has a significant computational cost, especially for graphs. 

Regarding the fusion of several methods, when focusing on text only, the best results are obtained with the hybrid combination of CamemBERT methods applied to word and message levels. The improvement is, however, very small compared to the methods used on their own. The early fusion between GraphWave (vertex-level) and WDA-SG2V (whole-graph level) is the best combination of graph embedding methods. Finally, the overall best performance is achieved by the fusion of text and graphs. The hybrid fusion of CamemBERT with the graph-based baseline achieves an 87.06~\% $F$-measure, with an 86.21~\% precision and an 87.93~\% recall. This result shows that, for the task of abuse detection, different modalities contain different types of information and that combining them brings real benefits compared to using them separately, thereby answering RQ2 (Does combining the two modalities considered in this paper help improving abuse detection performance?). 

\begin{table}[htb!]
    \centering
    \begin{tabular}{l l r r r}
        \toprule
        \textbf{Type} & \textbf{Methods} & \textbf{Dim.} & \textbf{$F$-measure} & \textcolor{highlightCol}{\textbf{Runtime}} \\
        \midrule
            Baseline & Baseline-Graph & 477 & 83.40 & \textcolor{highlightCol}{3:51:25} \\
            & Baseline-Text & 29 & 75.21 & \textcolor{highlightCol}{0:41} \\
        \midrule
            Text embedding & CamemBERT-W & 1,024 & 81.02& \textcolor{highlightCol}{6:56} \\
        \midrule
            Graph embedding & WDA-SG2V & 128 & 80.61 & \textcolor{highlightCol}{7:02} \\
        \midrule
            Fusion text-text & CamemBERT-W + CamemBERT-M & 2,048 & 81.32 & \textcolor{highlightCol}{14:37} \\
        \midrule
            Fusion graph-graph & WDA-SG2V + GraphWave & 256 & 82.57 & \textcolor{highlightCol}{25:31} \\
        \midrule
            Fusion graph-text & Baseline-Graph + CamemBERT-W & 1,501 & \textbf{87.06} & \textcolor{highlightCol}{4:02:56} \\
        \botrule
    \end{tabular}
    \caption{Summary of the best-performing method in terms of $F$-measure for each type of approach studied in this article. Column \textit{Runtime} indicates the total runtime (\textit{hour:minute:seconds}).}
    \label{tab:EmbeddingsBestPerfs}
\end{table}

Regarding the two graph embedding methods that we propose in this paper (WDA-SG2V and WDA-WSGCN), both obtain better results than their original counterpart, which do not use all the vertex and edge attributes. This shows that integrating such information directly in the representation learning process is important and allows learning more informative embeddings.

%%%%%%%%%%%%%%%%%%%%%%%%%%%%%%%%%%%%%%%%%%%%%
\section{Feature Study}
\label{sec:EmbeddingFeatureStudy}
In the previous section, we showed that textual embeddings allow improving the detection of abusive comments over a feature-based baseline, while graph embeddings are slightly less effective but a lot faster than the selected graph-based features. \textcolor{highlightCol}{However, none of the vector-based representations generated by these automated methods are inherently interpretable in terms of classification decisions. To better understand the properties captured or overlooked by these representation methods, we propose an analysis of both text and graph embedding approaches.}

In~\cite{Papegnies2019, Cecillon2024phd}, the authors propose methods based on text and graph features that we use as baselines in Section~\ref{sec:ClassificationResults}. For each method, they identify the most discriminative features for the classification task, which they call \textit{Best Features} (BF). They define subsets of 3 features for the text-based approach and 10 features for the graph-based approach, which are sufficient to reach 97~\% of the original performance (obtained when considering the complete feature set). 

Our analysis aims at determining whether these \textit{Best Features} are efficiently captured by the embeddings. To this end, we compare the $F$-measure score obtained by each embedding method on its own, with the score obtained by using an additional dimension integrating one of the best features.
\begin{itemize}
    \item If the performance significantly increases, we conclude that the considered Best Feature is not captured by the embedding. These cases are represented in red in Figures~\ref{fig:EmbeddingTextAblation} and~\ref{fig:EmbeddingGraphAblation}.
    \item If the performance stays the same or increases by less than 0.50 point, we conclude that the structural property corresponding to the considered Best Feature is well captured by the embedding (shown in green).
    \item If the performance increase is higher than 0.50 point but not statistically significant, we conclude that the considered Best Feature is only partially captured by the embedding method (represented in orange).
\end{itemize}
We first study text-related features (Section~\ref{sec:EmbeddingFeatureStudyText}), then graph-related measures (Section~\ref{sec:EmbeddingFeatureStudyGraph}).

%%%%%%%%%%%%%%%%%%%
\subsection{Text Features}
\label{sec:EmbeddingFeatureStudyText}
\begin{figure}[htb!]
    \center
    \includegraphics[width=0.66\textwidth]{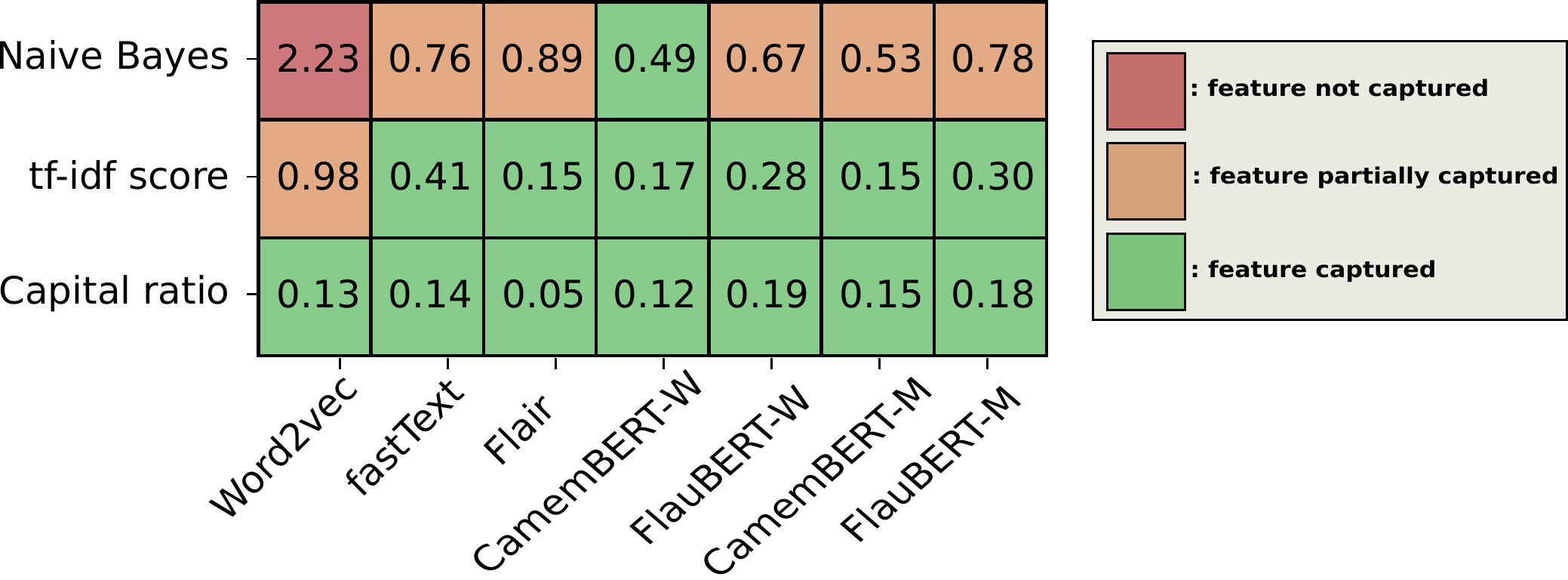}
    \caption{Text measures captured (green), partially captured (orange), or not captured (red) by the word embedding approaches. Each value is the difference between the $F$-measure score obtained by the embedding method on its own, and the score obtained by the embedding method complemented by the corresponding Best Feature.}
    \label{fig:EmbeddingTextAblation}
\end{figure}

Results for text best-features addition are shown in Figure~\ref{fig:EmbeddingTextAblation}. The ratio of capital letters in the message seems to be well captured by all the embedding methods. The same observation applies to the TF-IDF score computed over the \textit{Abuse} class except for Word2vec, which only partially captures this measure. In contrast, no method completely captures the information conveyed by the \textit{Naive Bayes} score. Word2vec is even completely unable to capture it. This feature comes from a fully-fledged classifier and is, by far, the most important one from the content-based baseline. Therefore, we can suppose that this information is too complex to be modeled by embedding methods. Furthermore, it integrates several types of information which might be difficult for an embedding to capture. However, almost all these methods yield a better $F$-measure than the baseline. Therefore, we can suppose that these methods might be able to capture other properties of the message that are not represented by any text-related feature, but improve the overall performance when all combined through the embedding. 
Another interesting observation of this study is that we can understand why Word2vec performs worse than the baseline and is the least efficient embedding method, as it fails to capture the two most important features of that baseline.

%%%%%%%%%%%%%%%%%%%
\subsection{Graph Features}
\label{sec:EmbeddingFeatureStudyGraph}
\begin{figure}[htb!]
    \center
    \includegraphics[width=0.8\textwidth]{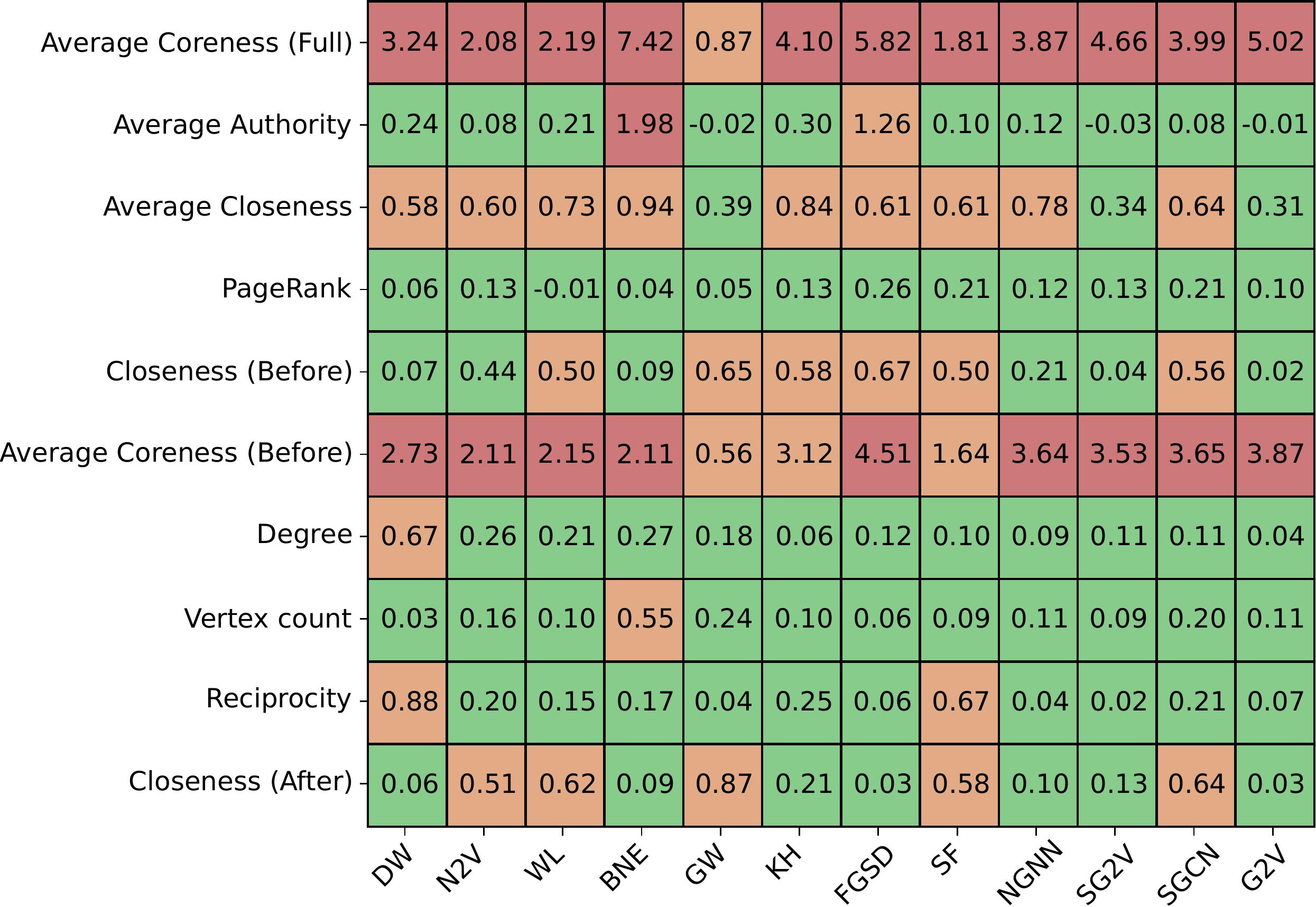}
    \caption{Topological measures captured (green), partially captured (orange), or not captured (red) by the embedding approaches. The first 4 topological measures are computed at the \textit{Graph} level and the last 5 topological measures are computed at the \textit{Vertex} level. Each value is the difference between the $F$-measure score obtained by the embedding method on its own and the score obtained by the embedding method complemented by the corresponding Best Feature.}
    \label{fig:EmbeddingGraphAblation}
\end{figure}

An interesting result shown by Figure~\ref{fig:EmbeddingGraphAblation} is that some topological measures seem to be well captured by almost all the embedding methods (e.g. Authority score at graph level, PageRank centrality, Degree centrality, Vertex count, and Reciprocity). Contrariwise, the Coreness score at graph level on the Full and Before graphs are partially captured by GraphWave and Spectral Features, and not captured at all by all other methods. The Closeness Centrality at the graph level and the vertex level on the Before and After graphs are well captured by some methods and partially captured by others. BoostNE fails to capture the authority at the graph level, which might explain why this method obtains the worst results among vertex embeddings.
Graph2vec, SG2V and our proposed method WDA-SG2V, 3 whole-graph embedding methods, are the only ones to correctly capture all the features except for the 2 mentioned above. There is no surprise that they are 3 of the best-performing methods, together with GraphWave. The latter, considered as a vertex embedding method, is the only approach to capture, at least partially, all the best features. However, these 4 methods obtain lower $F$-measure scores than the baseline. We assume that they might not capture other properties of the graph which are less important but improve the performance when all combined. 

An interesting result of this study is that there is no clear difference in the type of information captured by vertex and whole-graph embedding approaches. Vertex embedding methods can capture certain graph-scale topological measures, and whole-graph embedding methods can capture some vertex-scale measures. This property may result from the relatively small size of our graphs, as the second-order neighborhood of a vertex might include the majority of the graph. Thus, differences between vertex and whole-graph embedding methods are not as important as they could be on larger graphs. Furthermore, our graphs are built around a specific vertex. This specificity might help the whole graph embeddings to capture better vertex-level information. 

\bigskip
The results presented in Sections~\ref{sec:EmbeddingFeatureStudyText} and~\ref{sec:EmbeddingFeatureStudyGraph} allow answering RQ3 (Which discriminant information do representation learning methods automatically capture or miss, compared to feature engineering?). It appears that the best text embedding method captures all the discriminant text features from the baseline, and probably even more information, considering its better performance. The best graph embedding methods fail to capture two of the most discriminant features from the graph baseline, which explain their slightly lower performance.

%%%%%%%%%%%%%%%%%%%%%%%%%%%%%%%%%%%%%%%%%%%%%
\section{Conclusion}
\label{sec:Embeddingconclusion}
In this article, we have experimented with embedding methods to automate the representation learning process as part of an automatic abuse detection task. Our experiments cover the textual and contextual aspects of that task. We first compared five text embedding methods to learn representations of messages in French. The contextualized embeddings were much more efficient than the fixed ones. The best approach, relying on CamemBERT, achieved up to 81.02~\% $F$-measure. This is a massive improvement over a baseline based on a manually crafted set of features. 
We then compared 13 graph embedding methods, half of them generating embeddings at the vertex scale, and the other half treating the graph as a whole. We proposed variants of the SG2V and WSGCN models~\cite{Cecillon2024} able to treat weighted, directed, and signed graphs with vertex attributes. Both variants obtain better results than the original models, showing that they convey important information for this task. The performance obtained in a language-independent setting, where edge signs are ignored, is slightly lower but remains comparable. In addition, we combined different types of embeddings and showed that using textual content and conversational structure allows taking advantage of both sources of information, thereby significantly improving classification performance. 

We think that these results are very promising in multiple ways. They highlight the importance of introducing edge and vertex attributes in the representation learning process for graphs. The learned graph embeddings are much more robust, and although they obtain slightly lower classification scores than the feature-engineered baseline, they still demonstrate great performance. This is particularly interesting since these methods are completely task-independent, much more scalable, and time-efficient. 
\textcolor{highlightCol}{We believe our work paves the way for new research directions. In this study, we independently learn text and graph embeddings before combining them. A promising avenue for future research would be to train directly in a multimodal way, i.e. to integrate these aspects during the representation learning process, producing a unified embedding that captures all relevant information. While existing methods have been designed to jointly train on both graph and text data~\cite{Brannon2023}, they do not fully exploit the rich information embedded in our conversational graphs (structure, attributes, directions, weights, signs). 
Another promising perspective is to include time as another modality: messages are posted in a specific order, forming a sequence. Conversations are therefore evolving objects, that can be modeled through dynamic networks. The field of representation learning for dynamic data is well-established~\cite{Barros2021, Zheng2024}, and integrating existing methods into our approach could further enhance abuse detection performance.} 

% \section*{Declarations}
% All authors certify that they have no affiliations with or involvement in any organization or entity with any financial interest or non-financial interest in the subject matter or materials discussed in this manuscript.

\bibliography{Cecillon2024.bib}% common bib file

%%%%%%%%%%%%%%%%%%%%%%%%%%%%%%%%%%%%%%%%%%%%%
\appendix
\section{\textcolor{highlightCol}{Additional Results}}
\label{sec:Appendix}
\textcolor{highlightCol}{In order to complement the classification results provided in Section~\ref{sec:ClassificationResults}, the following tables provide the Precision and Recall scores used to compute the $F$-measure scores shown in the main text. Table~\ref{tab:EmbeddingPerfLexicalPrecRec} corresponds to Table~\ref{tab:EmbeddingPerfLexical}; Table~\ref{tab:EmbeddingPerfGraphPrecRec} to Table~\ref{tab:EmbeddingPerfGraph}; and Table~\ref{tab:EmbeddingFusionPrecRec} to Table~\ref{tab:EmbeddingFusion}.} 

\begin{table}[htb!]
    \color{highlightCol}
    \centering
    \begin{tabular}{l l r r r r }
        \toprule
        \textbf{Scale} & \textbf{Method} & \textbf{Dimension} & \textbf{Precision} & \textbf{Recall} & \textbf{Total Runtime} \\
        \midrule
            -- & Baseline-Text & 29 & 77.36 & 73.17 & 0:41\\
        \midrule
            Word & Word2vec & 300 & 72.59 & 69.74 & 4:14\\
            & fastText & 300 & 75.20 & 76.83 & 4:29\\
            & Flair & 2,048 & 77.16 & 77.50 & 5:25 \\
            & CamemBERT-W & 1,024 & 82.69 & 79.41 & 6:56\\
            & FlauBERT-W & 1,024 & 80.64 & 76.39 & 7:11\\
        \midrule
            Message & CamemBERT-M & 1,024 & 80.01 & 81.93 & 4:59\\
            & FlauBERT-M & 1,024 & 75.44 & 74.54 & 5:08\\ 
        \botrule
    \end{tabular}
    \caption{Precision and Recall performance obtained by the baseline approach (based on feature engineering) and the five word embedding methods described in Section~\ref{sec:EmbLexicalMethods}. The runtime is expressed as \textit{minutes:seconds}. The corresponding $F$-measure scores are provided in Table~\ref{tab:EmbeddingPerfLexical}.}
    \label{tab:EmbeddingPerfLexicalPrecRec}
\end{table}

\begin{table}[htb!]
    \color{highlightCol}
    \centering
    \begin{tabular}{l l r r r r}
        \toprule
        \textbf{Scale} & \textbf{Method} & \textbf{Dimension} & \textbf{Precision} & \textbf{Recall} & \textbf{Total Runtime}\\[2pt]
        \midrule
           -- & Baseline-Graph & 477 & 81.08 & 85.85 & 3:51:25 \\
        \midrule
            Targeted vertex & DeepWalk & 128 & 75.68 & 71.85 & 14:29 \\
            & Node2vec & 128 & 76.44 & 72.82 & 15:01 \\
            & Walklets & 128 & 77.21 & 72.61 & 16:21 \\
            & BoostNE & 136 & 70.56 & 72.52 & 16:00\\
            & GraphWave & 200 & 81.64 & 79.08 & 15:12 \\
            & $k$-hop GNN & 64 & 77.85 & 70.44 & 24:55 \\
        \midrule
            Whole graph & FGSD & 200 & 72.64 & 69.68 & 6:35 \\
            & Spectral Features & 128 & 75.55 & 72.95 & 6:57\\
            & NGNN & 64 & 75.10 & 73.92 & 16:58 \\
            & Graphormer & 768 & 72.13 & 75.69 & 14:11 \\
            & Graph2vec & 128 & 74.99 & 81.63 & 4:54 \\
            & SG2V & 128 & 79.24 & 79.02 & 6:24 \\ 
            & WDA-SG2V & 128 & 82.33 & 78.96 & 7:02\\ 
            & WSGCN & 128 & 77.64 & 72.04 & 14:14 \\
            & WDA-WSGCN & 128 & 76.76 & 74.86 & 15:07 \\ [2pt]
        \botrule
    \end{tabular}
    \caption{Precision and Recall performance of the baseline and graph embedding methods, including the 13 selected in Section~\ref{sec:EmbGraphMethods}, and the two proposed in Section~\ref{sec:ProposedMethods}. The runtime is expressed as \textit{hour:minute:seconds}. The corresponding $F$-measure scores are provided in Table~\ref{tab:EmbeddingPerfGraph}.}
    \label{tab:EmbeddingPerfGraphPrecRec}
\end{table}

\begin{table}[htb!]
    \color{highlightCol}
    \centering
    \begin{tabular}{l l r r r r}
        \toprule
        \textbf{Fusion} & \textbf{Methods} & \textbf{Dim.} & \textbf{Precision} & \textbf{Recall} & \textbf{Runtime} \\
        \midrule
            Early & Graph Base + Text Base. & 506 & 83.05 & 86.02 & 3:55:17 \\
        \cline{2-6}
            & CamemBERT-W + CamemBERT-M & 2,048 & 82.71 & 79.97 & 14:37 \\
            & CamemBERT-W + FlauBERT-W & 2,048 & 85.73 & 74.72 & 16:12 \\
        \cline{2-6}
            & WDA-SG2V + GraphWave & 256 & 85.28 & 80.03 & 25:31 \\
        \cline{2-6}
            & WDA-SG2V + CamemBERT-W & 1,152 & 85.17 & 86.52 & 17:21 \\
            & GraphWave + CamemBERT-W & 1,152 & 82.83 & 89.39 & 24:59 \\
            & Graph Base. + CamemBERT-W & 1,501 & 85.14 & 88.46 & 4:01:26 \\ 
            & WDA-SG2V + Text Base & 157 & 79.95 & 84.98 & 9:47 \\ 
        \midrule
            Late & Graph Base. + Text Base. & 2 & 80.05 & 88.73 & 3:54:57 \\
        \cline{2-6}
            & CamemBERT-W + CamemBERT-M & 2 & 79.51 & 80.92 & 14:24 \\
            & CamemBERT-W + FlauBERT-W & 2 & 80.87 & 79.34 & 16:03 \\
        \cline{2-6}
            & WDA-SG2V + GraphWave & 2 & 84.18 & 80.29 & 25:32 \\
        \cline{2-6}
            & WDA-SG2V + CamemBERT-W & 2 & 87.81 & 82.91 & 16:29 \\
            & GraphWave + CamemBERT-W & 2 & 86.11 & 85.49 & 24:46 \\
            & Graph Base. + CamemBERT-W & 2 & 84.88 & 87.58 & 4:00:23 \\ 
            & WDA-SG2V + Text Base & 2 & 83.34 & 81.22 & 9:33 \\ 
        \midrule
            Hybrid & Graph Base. + Text Base. & 508 & 83.23 & 86.78 & 3:57:44 \\
        \cline{2-6}
            & CamemBERT-W + CamemBERT-M & 2,050 & 78.42 & 84.11 & 15:02 \\
            & CamemBERT-W + FlauBERT-W & 2,050 & 82.74 & 78.07 & 16:59 \\
        \cline{2-6}
            & WDA-SG2V + GraphWave & 258 & 81.61 & 81.07 & 27:44 \\
        \cline{2-6}
            & WDA-SG2V + CamemBERT-W & 1,154 & 83.64 & 89.07 & 19:17 \\
            & GraphWave + CamemBERT-W & 1,154 & 83.48 & 88.91 & 27:38 \\
            & Graph Base. + CamemBERT-W & 1,503 & 84.41 & 89.87 & 4:02:56\\ 
            & WDA-SG2V + Text Base & 159 & 81.93 & 83.18 & 10:25 \\
        \botrule
    \end{tabular}
    \caption{Precision and Recall obtained by the fusion of the best performing embedding methods following the three fusion strategies (Early, Late, Hybrid). Column \textit{Dim.} denotes the dimension of the vector representation, and \textit{Runtime} is the total runtime. The table is split horizontally into three parts, each one dedicated to one fusion strategy. In each such part, the same methods are gathered in four groups, depending on the nature of the combined methods. From top to bottom: baselines, fusion of text embeddings, fusion of graph embeddings, and fusion of text and graph embeddings. The corresponding $F$-measure scores are provided in Table~\ref{tab:EmbeddingFusion}.}
    \label{tab:EmbeddingFusionPrecRec}
\end{table}

\end{document}

%% file: figures/TunedG2Vauthor.tex
\begin{tikzpicture}

    \node [nn,fill=red!50,minimum size=1cm] (v1) at ( 6.00,  0.00) {$id_{74}$};
    \node [nn,minimum size=1cm] (v2) at ( 7.00, -1.00) {$id_{07}$};
    \node [nn,minimum size=1cm] (v3) at ( 8.50,  -1.25) {$id_{28}$};
    \node [nn,minimum size=1cm] (v4) at ( 9.50,  0.00) {$id_{99}$};
    \node [nn,minimum size=1cm] (v5) at ( 8.25,  1.50) {$id_{34}$};
    \node [nn,minimum size=1cm] (v6) at ( 6.75,  1.50) {$id_{57}$};

    \draw (v1) edge (v2);
    \draw (v1) edge (v4);
    \draw (v1) edge (v6);
    \draw (v2) edge (v3);
    \draw (v2) edge (v4);
    \draw (v4) edge (v5);
    \draw (v4) edge (v6);
    \draw (v5) edge (v6);

    \node at (7.50,-1.75) {$a)$};
\end{tikzpicture}

%% file: figures/TunedG2Vdistance.tex
\begin{tikzpicture}

    \node [nn,fill=red!50,minimum size=1cm] (v1) at ( 6.00,  0.00) {$0$};
    \node [nn,minimum size=1cm] (v2) at ( 7.00, -1.00) {$1$};
    \node [nn,minimum size=1cm] (v3) at ( 8.50,  -1.25) {$2$};
    \node [nn,minimum size=1cm] (v4) at ( 9.50,  0.00) {$1$};
    \node [nn,minimum size=1cm] (v5) at ( 8.25,  1.50) {$2$};
    \node [nn,minimum size=1cm] (v6) at ( 6.75,  1.50) {$1$};

    \draw (v1) edge (v2);
    \draw (v1) edge (v4);
    \draw (v1) edge (v6);
    \draw (v2) edge (v3);
    \draw (v2) edge (v4);
    \draw (v4) edge (v5);
    \draw (v4) edge (v6);
    \draw (v5) edge (v6);

    \node at (7.50,-1.75) {$b)$};
\end{tikzpicture}

%% file: figures/TunedG2Vtarget.tex
\begin{tikzpicture}

    \node [nn,fill=red!50,minimum size=1cm] (v1) at ( 6.00,  0.00) {$1$};
    \node [nn,minimum size=1cm] (v2) at ( 7.00, -1.00) {$0$};
    \node [nn,minimum size=1cm] (v3) at ( 8.50,  -1.25) {$0$};
    \node [nn,minimum size=1cm] (v4) at ( 9.50,  0.00) {$0$};
    \node [nn,minimum size=1cm] (v5) at ( 8.25,  1.50) {$0$};
    \node [nn,minimum size=1cm] (v6) at ( 6.75,  1.50) {$0$};

    \draw (v1) edge (v2);
    \draw (v1) edge (v4);
    \draw (v1) edge (v6);
    \draw (v2) edge (v3);
    \draw (v2) edge (v4);
    \draw (v4) edge (v5);
    \draw (v4) edge (v6);
    \draw (v5) edge (v6);

    \node at (7.50,-1.75) {$c)$};
\end{tikzpicture}